\documentclass[
amssymb,floatfix,floats,
aps,prl,preprintnumbers,noshowpacs,twocolumn,superscriptaddress]{revtex4-1}

\usepackage{booktabs}
\usepackage{color,graphicx}
\usepackage[dvipsnames]{xcolor}
\usepackage{dcolumn}
\usepackage{upgreek}
\usepackage{bm}
\usepackage[hidelinks,colorlinks=true,allcolors=blue,citecolor=blue]{hyperref}
\usepackage{chngcntr}
\usepackage[thinspace,thinqspace]{SIunits}
\usepackage{amsfonts}
\usepackage[ulem=normalem]{changes}
\usepackage{orcidlink}
\usepackage{lineno}

\newcommand{\jld}{\ensuremath{J_\text{1D}}}
\newcommand{\D}{\ensuremath{\Delta}}
\newcommand{\jEd}{\ensuremath{J_\text{3D}}}
\newcommand{\Bl}{\ensuremath{B_\text{c1}}}
\newcommand{\Bu}{\ensuremath{B_\text{c2}}}

\newcommand{\rs}[1]{\textcolor{black}{#1}}

\newcommand{\red}[1]{\rs{\sout{}}}

\begin{document}
\title{Magnetic-field-induced Tomonaga-Luttinger liquid and Bose-Einstein condensate phases in an organic $S=1$ Haldane chain}

\author{Ivan Jakovac\,\orcidlink{0009-0005-0393-3601}}
\affiliation{Department of Physics, Faculty of Science, University of Zagreb, Bijenička cesta 32, 10000 Zagreb, Croatia}

\author{Mihael S. Grbi{\'c}\,\orcidlink{0000-0002-2542-2192}}
\email{corresponding author: mgrbic.phy@pmf.hr}
\affiliation{Department of Physics, Faculty of Science, University of Zagreb, Bijenička cesta 32, 10000 Zagreb, Croatia}

\author{Maxime Dupont\,\orcidlink{0000-0001-5719-5687}}
\affiliation{Laboratoire de Physique Théorique, Université Paul Sabatier, 118 Route de Narbonne, 31062 Toulouse Cedex 4, France}
\affiliation{Present address: Rigetti Computing, 775 Heinz Avenue, Berkeley, California 94710, USA}

\author{Nicolas Laflorencie\,\orcidlink{0000-0001-8634-1006}}
\affiliation{Laboratoire de Physique Théorique, Université Paul Sabatier, 118 Route de Narbonne, 31062 Toulouse Cedex 4, France}

\author{Sylvain Capponi\,\orcidlink{0000-0001-9172-049X}}
\affiliation{Laboratoire de Physique Théorique, Université Paul Sabatier, 118 Route de Narbonne, 31062 Toulouse Cedex 4, France}

\author{Yuko Hosokoshi\,\orcidlink{0000-0002-4480-4137}}
\affiliation{Department of Physics, Osaka Metropolitan University, Osaka 599-8531, Japan}
\affiliation{Department of Physical Science, Osaka Prefecture University, Osaka 599-8531, Japan}

\author{Steffen Kr{\"a}mer\,\orcidlink{0000-0002-6107-3583}}
\affiliation{Laboratoire National des Champs Magn\'{e}tiques Intenses, LNCMI-CNRS (UPR3228), EMFL, \\ Univ.  Grenoble Alpes, Univ. Toulouse, INSA-T, 38042 Grenoble Cedex 9, France}

\author{Yurii Skourski\,\orcidlink{0000-0002-4100-6420}}
\affiliation{Hochfeld-Magnetlabor Dresden (HLD-EMFL) Helmholtz-Zentrum Dresden-Rossendorf, 01328 Dresden, Germany}

\author{Sven Luther}
\affiliation{Hochfeld-Magnetlabor Dresden (HLD-EMFL) Helmholtz-Zentrum Dresden-Rossendorf, 01328 Dresden, Germany}

\author{Masashi Takigawa\,\orcidlink{0000-0002-4516-5074}}
\affiliation{Institute for Solid State Physics, University of Tokyo, 5-1-5 Kashiwanoha, Kashiwa, Chiba 277-8581, Japan}

\author{Mladen Horvati{\'c}\,\orcidlink{0000-0001-7161-0488}}
\affiliation{Laboratoire National des Champs Magn\'{e}tiques Intenses, LNCMI-CNRS (UPR3228), EMFL, \\ Univ.  Grenoble Alpes, Univ.  Toulouse, INSA-T, 38042 Grenoble Cedex 9, France}

\date{\today} 
\begin{abstract}
We present a comprehensive low-temperature $^1$H nuclear magnetic resonance (NMR) study of magnetic field-induced phases and spin fluctuations in an isotropic Haldane chain system $m$-NO$_2$PhBNO (BoNO) in fields up to 34~T. NMR spectral and $T_1^{-1}$ relaxation rate data are used to map out the transition temperature $T_c(B)$ and we show that it defines a complete phase boundary between the field-induced Bose-Einstein condensate (BEC) and Tomonaga-Luttinger liquid (TLL) phases by reproducing its shape via quantum Monte Carlo calculations. We also observe the theoretical critical behavior $T_c (B) \propto (B-\Bu)^{\nu}$ with $\nu = 2/3$ and universal quasiparticle scaling in the low-temperature limit close to \Bu. The extensive Bayesian inference analysis of the temperature and field dependence of $T_1 ^{-1} (T,B)$ reveals attractive interactions across the TLL phase of a Haldane chain. Our results are further corroborated by magnetostriction measurements in pulsed magnetic fields and density matrix renormalization group calculations.
\end{abstract}
\maketitle

\begin{figure}[ht!]
	\centering
	\includegraphics[width=0.5\textwidth]{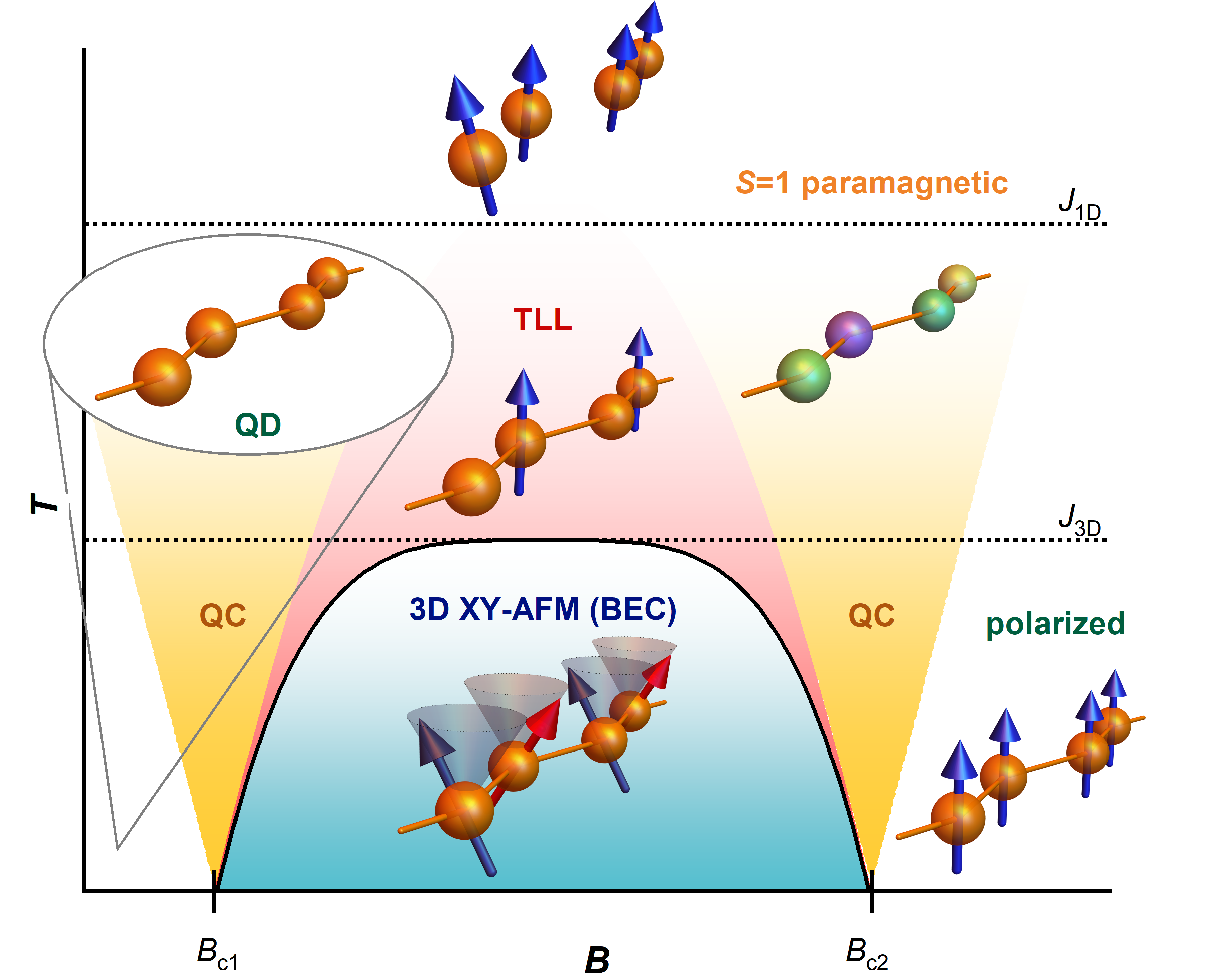}
	\caption{
		In the absence of the external magnetic field, a gap separates the singlet quantum-disordered (QD) ground state from the first excited triplet. The energy of the triplet is lowered via the Zeeman effect in a finite magnetic field, eventually closing the gap at \Bl. In the range \Bl $< B <$ \Bu, the system is gapless and can be described by the TLL framework below energy scale set by \jld. At lower temperatures, \jEd\ becomes relevant, and the system enters a long-range coherent 3D XY-AFM state, corresponding to the BEC in the bosonic mapping. Crossovers are indicated by dotted lines, while the solid line marks the phase transition. Universal quasiparticle behavior is observed in the quantum-critical (QC) region. At higher temperatures, system is an isotropic 3D $S=1$ paramagnet.}
	\label{fig:fig1}
\end{figure}

Low-dimensional quantum magnets -- small spin ($S$\,$=$\,$1/2$ or $S$\,$=$\,$1$) chain or ladder networks where quantum fluctuations control the ground state – allow development of theoretical models directly comparable to experiment. This is essential for understanding quantum phenomena such as the Tomonaga-Luttinger liquid (TLL)~\cite{GiamarchiBook} and long range order (LRO) of Bose-Einstein condensate (BEC) type~\cite{Zapf2014}. The TLL theory provides a universal description of gapless systems with interacting quasiparticles~\cite{Klanjsek2008,Jeong2013,Vijayan,Wu2019} in one dimension (1D), extending to topological matter~\cite{Bouchoule2025} and quantum information devices~\cite{Keimer2017}. In 1D spin systems, low-energy excitation spectra can be modelled as quasiparticles determined by two parameters: renormalized Fermi velocity $u$ and TLL interaction parameter $K$. At low temperatures, finite three-dimensional coupling (\jEd) between 1D units drives a 3D XY antiferromagnetic (AFM) order -- a BEC of a magnetic spin system mapped onto a hard-core boson model. In the studied TLL and BEC phases, magnetic field acts as a tunable chemical potential, making quantum magnets a perfect testing ground for theoretical predictions~\cite{Klanjsek2008}, study of quantum phenomena~\cite{Autti,Aveline,Sompet2022}, and an ideal platform for quantum computing~\cite{Byrnes,Andrianov,Mohseni2022}.
\\
\indent Arguably, the most remarkable spin chain example is the $S=1$ Haldane chain. Over 40 years ago Haldane discovered~\cite{HaldanePRL} that a 1D system of integer spins described by the Heisenberg antiferromagnetic ($J_\text{1D}>0$) Hamiltonian $H_0 = J_\text{1D} \Sigma_{ij} \mathbf{S}_i \mathbf{S}_j$ hosts a non-magnetic singlet ground state separated by an energy gap \D\ from the first excited triplet state. This contrasts half-integer spin chains, where the ground state is gapless, implying that \D\ emerges from a quantum nature of the system. The Haldane phase remains stable~\cite{Sakai1990,WierschemPRL} under relatively small single-ion anisotropy ($D$) and finite \jEd, inevitable in any real material. In an external field ($B$), Zeeman splitting lowers the energy of the $S_z=+1$ triplet band, closing the gap at $\Bl = \Delta / g\mu_B$ (Fig.~\ref{fig:fig1}). Therefore, \Bl\ is a quantum-critical point (QCP), as quantum fluctuations drive the transition at $T=0~$K. For $B \geq \Bl$, the spectrum is gapless and for $T\le \jld$ the TLL state with attractive interactions emerges. Further increase of $B$ fills the fermion band and tunes the TLL parameter $K$. Due to the \jEd\ interactions, a 3D long-range ordered BEC forms below $T_c(\jEd)$. Both phases persist up to $\Bu \approx 4 \jld / g\mu_B$, where the system fully polarizes and the spectrum is again gapped. This rich phase diagram motivated extensive research, and more than a dozen Haldane-phase compounds have been synthesized~\cite{Maximova}. Remarkably, none allowed a comprehensive study of TLL and BEC phases across the $B-T$ diagram (see Supplemental Material (SM)~\cite{Supp} for details and references~\cite{Broholm2002,Maximova,DARRIET1993,Xu,Takeuchi92,Zheludev96,Takeuchi92b,Meyer,Honda,Koike,Johnson1979,Buyers1986,Morra,Brambley,Lipps2017,Vasiliev,Law2014,Christian,Williams2020,Zheludev2000,Smirnov2002,Tsujii2005b,Pahari,Bera,TinNatComm,Cizmar2008,Li2025,Chou,GotoPRB,Koido,Cizmar2002}). In all systems the $S=1$ resides on a transition-metal ion, e.g. Ni$^{2+}$ in CsNiCl$_3$, Ni(C$_5$D$_{14}$N$_2$)$_2$N$_3$PF$_6$ (NDMAP), Ni(C$_2$H$_8$N$_2$)$_2$NO$_2$ClO$_4$ (NENP) and Ni(C$_3$H$_{10}$N$_2$)$_2$NO$_2$ClO$_4$ (NINO), where spin-orbit coupling adds Hamiltonian terms breaking the U(1) symmetry needed for BEC. In NENP, NINO and NDMAP, a field along the chain creates a staggered component, since the $g$-tensor's principal axes tilt by $\pm \theta$, that cannot be compensated by sample rotation. NINO and NENP also have finite axial ($D$) and rhombic ($E$) anisotropy, and CsNiCl$_3$ orders at zero field. In many systems \jld\ was too large to reach \Bu, or even \Bl. Consequently, most of the long-standing theoretical predictions on TLL~\cite{Konik2002,Fath03} or BEC~\cite{Affleck1990,Affleck1991} phases in a model Haldane chain remain unconfirmed.
\\
\indent Recently, a different route to Haldane $S=1$ chains has been demonstrated~\cite{Jakovac2024}. The fully organic nitroxide radical compound 3,5-bis(N-tert-butylaminoxyl)-3'-nitrobiphenyl ($m$-NO$_2$PhBNO, BoNO) contains $S=1$ units formed by two ferromagnetically coupled ($|J_\text{FM}| \gtrsim 500$~K) aminoxyl radicals. The delocalized spin density (Fig.~S1 in the SM~\cite{Supp}) strongly suppresses spin-orbit coupling, preserving the symmetry of the spin Hamiltonian. BoNO crystallizes in an orthorhombic \textit{Pbca} structure with chains along the $a$ axis and $z=6$ neighboring chains forming a distorted hexagonal lattice (Fig.~S2 in the SM~\cite{Supp}). The nearly isotropic $g$ tensor ($g_a=2.0041$, $g_b=2.0065$, $g_c=2.0063$) is close to the free-electron value $g_e=2.0023$~\cite{Jakovac2024}. Low-field susceptibility is consistent with an antiferromagnetic $S=1$ chain with \jld~$=(11.1\pm0.2)$~K and total interchain coupling $z$\jEd~$\approx0.6$~K. Magnetization at 1.3~K increases quasi-linearly between \Bl~$\approx2$~T and \Bu~$\approx33$~T, consistent with coupled $S=1$ chains with $z$\jEd/\jld=0.04. BoNO is therefore a model Haldane-chain material whose accessible critical fields enable mapping of the complete field-induced $B$--$T$ phase diagram.
\\
\indent In this Letter we present a $^1$H nuclear magnetic resonance (NMR) study of magnetic-field induced phases and spin fluctuations in BoNO, in high magnetic fields (up to 34~T) and at temperatures down to 460~mK. To achieve this, we used an advanced NMR setup which allows measurements at high frequencies up to 1440~MHz. Our data reveal the long-sought evidence of attractive interactions across the TLL phase of a Haldane chain. Furthermore, we map out the field-induced BEC transition temperature $T_c(B)$ and reproduce its shape via quantum Monte Carlo (QMC) calculations. We verify the theoretical prediction $T_c (B) \propto (B-\Bu)^{\nu}$ with $\nu = 2/3$~\cite{Nikuni} in the low-$T$ limit close to \Bu. Our results are corroborated by magnetostriction measurements in pulsed magnetic fields, density matrix renormalization group (DMRG) calculation and extensive Bayesian inference analysis.
\begin{figure*}[ht!]
	\centering
	\includegraphics[width=\textwidth]{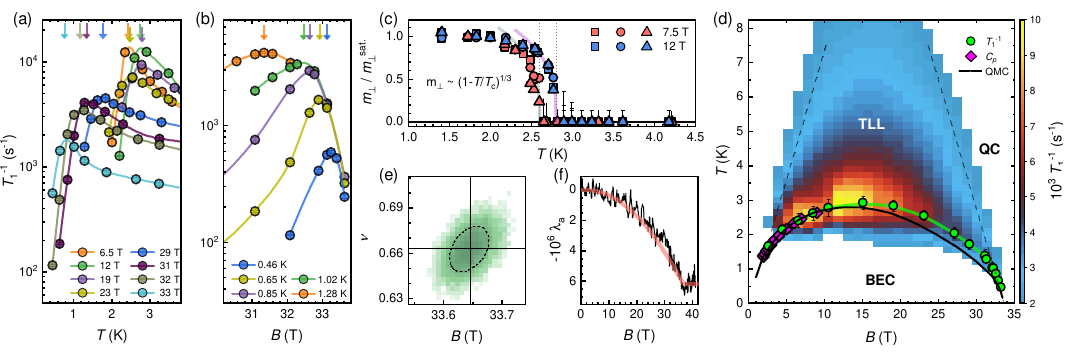}
	\caption{
		(a) Temperature and (b) magnetic-field dependence of the $^1$H relaxation rate $T_{1}^{-1}$ across the phase transition at various external magnetic fields and temperatures, respectively. Solid lines are cubic-spline fits, and arrows mark the estimated transition temperatures $T_c(B)$. (c) Temperature evolution of the ordered transverse magnetic moment $m_\perp \propto \Delta f$ across $T_c$ at $7.5$~T (red) and $12$~T (blue), determined from spectral splitting at various $^1$H NMR lines (see Figs.~S4 and S5 in the SM~\cite{Supp}). (d) Complete phase diagram derived from $T_{1}^{-1}$ and spectral data (green circles). At low magnetic fields, additional heat-capacity measurements (magenta diamonds) corroborate the NMR-determined phase boundary. QMC simulations (black line) reproduce the experimental phase diagram perfectly up to 15 T, above which a small discrepancy develops, which can be reconciled by allowing for a field-dependent interchain coupling $J_\text{3D}$. (e) Probability density map of the upper critical field \Bu\ and critical exponent $\nu$, calculated for $T_c \propto (\Bu-B)^{\nu}$ with the mean values of $\Bu=33.645~$T and $\nu=0.663$. Dashed line marks the 68\% highest density interval (HDI). (f) Measured magnetostriction along the chains $\lambda_a=(\Delta L_a/L_a)$ vs. $B$ at 1.4~K shows a parabolic contraction up to \Bu. Red line is a guide to the eye.}
	\label{fig:fig2}
\end{figure*}\\
\indent To investigate the properties of magnetic-field induced phases in BoNO for $B>$~\Bl\, we probe the electronic spin excitations via $^1$H NMR. The typical $^1$H spectrum of BoNO is defined by the Zeeman interaction via $f={^1}\gamma (1+\textbf{K}_s)\textbf{B}\textbf{I}$, where ${^1}\gamma/2\pi =42.5775$~ MHz/T is the hydrogen gyromagnetic constant, $\textbf{B}$ is the magnetic field, $\textbf{I}$ nuclear spin, and $\textbf{K}_s$ the nuclear shift tensor. The spectrum consists of a large number of spectral lines (see Fig.~S3 in the SM~\cite{Supp}) concentrated close to the bare nucleus Larmor frequency value $f_0 = {^1}\gamma B$ and several $^1$H spectral lines shifted away from $f_0$ due to large hyperfine coupling $\mathbb{A} \propto K_s $ to electron magnetic moments. The shifted lines correspond to~\cite{Jakovac2024} the hydrogen sites H$_1$, H$_3$ and H$_5$ on the aromatic, electron-rich part of the BoNO molecule, making them very sensitive probes of the electronic dynamics. For an arbitrary magnetic field orientation one expects 4 spectral lines for each $^1$H site, although due to the high crystal symmetry and significant spectral overlap fewer lines can be discerned.\\
\indent In Fig.~\ref{fig:fig2}(a) we show the relaxation rate $T_1 ^{-1} (T)$ dependence for various magnetic field values. Close to the saturation field, we did several isothermal $T_1 ^{-1}(B)$ measurements (Fig.~\ref{fig:fig2}(b)) at lowest temperatures. The data were collected on the isolated NMR line with the largest hyperfine shift to avoid spurious effects due to spectral line overlap. The $T_1 ^{-1}$ data probe the local electronic correlations in the low energy limit, making it a stringent probe of the system's ground state. For temperatures below $10$~K ($\approx\jld$), $T_1 ^{-1}(T)$ shows a slowly divergent behaviour as the system approaches the magnetic transition temperature $T_c$, where $T_1 ^{-1}$ reaches a maximum value. Across $T_c$, the $^1$H NMR spectra splits (Fig.~\ref{fig:fig2}(c) and Fig.~S4 in the SM~\cite{Supp}) due to the onset of transverse magnetisation $m_{\perp}$. The line splitting $\Delta f$ can be well described by the temperature dependence $(1-T/T_c)^{\beta}$, with an exponent $\beta \sim 1/3$ fully consistent with the 3D-XY universality class $\beta=0.3485(2)$~\cite{Campostrini}. Below $T_c$, critical fluctuations increase and $T_1 ^{-1}(T)$ show a steep decrease, similar to other BEC systems~\cite{Jeong2017,Blosser}. From the onset of splitting and the maximum in $T_1 ^{-1}$ data, we determine the $T_c (B)$ phase boundary of the 3D LRO (Fig.~\ref{fig:fig2}(d)).
\begin{figure*}[ht!]
	\centering
	\includegraphics[width=\textwidth]{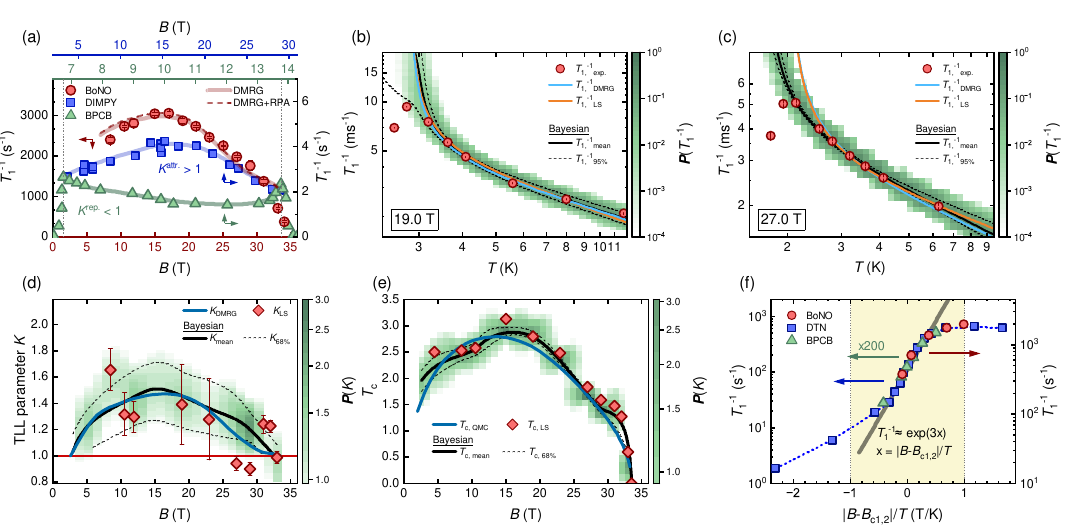}
	\caption{
		(a) Constant-temperature cuts of $T_1^{-1}$ in the TLL regime of BoNO (red circles, bottom and left axes), compared to those of ladder systems DIMPY (blue squares, top and right axes) and BPCB (green triangles,top and right axes) from Ref.~[\onlinecite{Jeong2016}]. The data on BoNO show attractive quasiparticle interactions ($K>1$), as expected for the Haldane system. Lines are plots of TLL expression~\ref{T1TLL} using the parameters calculated by DMRG (full lines), and for BoNO with additional RPA correction (dashed lines). (b, c) Experimental $T_1^{-1}(T)$ data (red circles) overlaid with Bayesian probability distributions (green colormap) at $19$~T and $27$~T. Bayesian mean (black full line) and 95\% HDI values (black dashed lines) are compared to the least-squares fits (LS, orange lines) and fits using theoretical values $K$ calculated by DMRG (blue lines).  (d, e) Probability density maps (green) with mean (black full lines) and 68\% HDI values (black dashed lines) for $K$ and $T_c$, compared with theory (blue lines) and LS fits (red diamonds). (f)~Universal quasiparticle scaling of $T_1^{-1}~(B)$ data measured across the QC region in BoNO (red circles), consistent with the three-magnon model (grey line), also observed~\cite{Mukhopadhyay2012} in DTN (blue squares) and BPCB (green triangles).}
	\label{fig:fig3}
\end{figure*} The boundary $T_c (B)$ has a slightly asymmetric shape, with the maximum value of $T_c ^{max.} = 2.9$~K  at $B \approx 15$~T. The large $T_c ^{max.} / \jld$ value implies a significant interchain coupling $z\jEd/\jld$, in agreement with the $z\jEd/\jld \approx 0.06$ value estimated from the magnetic susceptibility measurements. At lower magnetic fields, hyperfine shifts are less pronounced, and the NMR spectra shrinks to the point where NMR lines start to overlap. Here, we have performed several specific heat capacity measurements~\cite{CvitanicPhD}  down to $1.3$~K to verify that $T_c(B)$ was correctly determined by NMR.  Although the additional data provide a more detailed view of the phase boundary close to \Bl, as lower temperatures were not available, we estimate that \Bl\ value is $(1.0 _{-0.2} ^{+0.5})~$T. For the measurements at the high-field end, temperatures down to $460$~mK were reached and we apply the low-temperature (LT) theoretical limit $T_c ^{LT} (B) \propto |B-\Bu|^{\nu}$ to extrapolate $\Bu=(33.65\pm0.04)$~T and $\nu = (0.66 \pm 0.07)$ using a standard windowing~\cite{Sebastian}  method (Fig.~S6 in the SM~\cite{Supp}) and Bayesian inference analysis (BIA) (Fig.~\ref{fig:fig2}(e)) (see Methods). BIA~\cite{VonToussaint} calculates conditional probabilities of observing specific values of parameters ($\Bu,\nu$) for the given observed dataset $T_c(B)$ and in expectation of a well-defined functional dependence (here, $T_c ^{LT} (B)$). It also provides point estimates and confidence intervals, and is more reliable in the case of limited or noisy datasets. The obtained exponent $\nu$ corresponds precisely to the universality class of a BEC quantum critical point, for which $\nu = z/D$~\cite{GiamarchiTsvelik,Nikuni}, where $D=3$ is the spatial dimensionality and $z = 2$ is the dynamical exponent associated with a quadratic triplon dispersion. With no anisotropy terms in the Hamiltonian, the upper critical field is then defined by a simple expression: $g \mu_B \Bu = 4 \jld + 2 z \jEd$. For consistency, using \Bu\ and \jld\ we obtain $z\jEd=0.6\pm0.2$~K, in agreement with susceptibility and $\Bl(\jEd)$ (as calculated by Ref.~\onlinecite{WierschemPRL}).\\
\indent The phase boundary $T_c (B)$ is also calculated using the state-of-the-art quantum Monte Carlo (QMC) simulations. For simplicity, the calculation is done on an $S=1$ chain with tetragonally coordinated \jEd. This does not affect the results as long as $z\jEd$ remains unaltered~\cite{Sakai1989,WierschemMPLB}. Taking $z\jEd ^\text{QMC}=0.676~$K, the calculated curve precisely follows the experimental data up to 10~T, with a small deviation above $15~$T. We present several options to address this discrepancy. One of them is the question of suitability of existing models. \jEd\ in BoNO is not isotropic  - three different pairs of $\jEd$ are distributed with a mirror symmetry along the $ab$ plane. Specific geometry of these \jEd\ interactions may modify the shape of $T_c (B)$ boundary. Unfortunately, this is not easy to take into account, since exact values of all \jEd\ interactions are difficult to determine, both theoretically and experimentally, and the introduced complexity makes numerical calculations unfeasible. The other cause of the discrepancy is the variation of  \jEd\ and \jld\ in the high-field region by magneto-elastic coupling. In a separate QMC calculation (see Fig.~S7 in the SM~\cite{Supp}), we found that a 50\% increase in the tetragonally symmetric \jEd\ can account for the experimental $T_c(B)$ boundary. Such a large variation is unusual and indicates that other parameters, e.g. \jld, should also change. By a simple scaling we can estimate that an increase in \jld\ up to 12\% can also reproduce $T_c(B)$ data. Variations in the $z\jld$($\jEd$) were observed in similar compounds~\cite{zapf2007,Nomura} where noticeable magnetostriction $\lambda_i=(\Delta L_i/L_i)$ $(i=a,b,c)$ occurs. By measuring $\lambda_a$ along the chain direction in pulse magnetic fields up to 41.2~T at 1.4~K,  we detected a response of $-7\cdot10^{-6}$ (Fig.~\ref{fig:fig2}(f)), while $|\lambda_b|$ perpendicular to the chains was $\le8\cdot 10^{-7}$, which is on the sensitivity limit of the method.  While the exact evaluation of $z\jld$($\jEd$) from the $\lambda_a(B)$ data is beyond the scope of this paper, we can conclude that  variations of $\jld$($z\jEd$) can account for the discrepancy.\\
\indent In the TLL state of a strictly 1D system, we expect~\cite{Konik2002} the transverse excitations originating at $q=\pi$ to be dominant, which results with a power-law~\cite{GiamarchiTsvelik} temperature dependence of the $T_1 ^{-1}$ relaxation rate~\cite{Klanjsek2008}:
{\begin{equation}
		\fontsize{8pt}{8pt}\selectfont
		T_{1, \text{TLL}} ^{-1} = \frac{\hbar \gamma^2 A_\perp ^2 A_0 ^x}{k_B u} \cos \left(\frac{\pi}{4 K}\right) B\left(\frac{1}{4 K},1- \frac{1}{2K}\right) \left(\frac{2 \pi T}{u}\right) ^{\frac{1}{2K}-1},
		\label{T1TLL}
\end{equation}}
where $\gamma$ is the nuclear gyromagnetic ratio, $A_\perp$ is the transverse hyperfine coupling constant, $A_0 ^x$ is the amplitude of the spin correlation function, and $B(x,y)=\Gamma(x) \cdot \Gamma(y)/ \Gamma(x+y)$. $K(B)$ dependence is defined by the properties of the original spin Hamiltonian. In the presence of sizeable 3D interactions ($z\jEd/\jld \gtrsim 0.01$) that increase $T_c$ and reduce the span of the 1D regime,  $T_1 ^{-1} (T)$ dependence is enhanced and warped~\cite{Jeong2013,Dupont2018,Horvatic2020} by critical spin fluctuations. To account for the 3D effects and determine the correct $K$ values, expression (\ref{T1TLL}) should be modified: $T_1 ^{-1} = T_{1, \text{TLL}} ^{-1} \Phi(K, T_c/T)$, where $\Phi(K, T_c/T)$ is an analytical correction factor based on random phase approximation (RPA) approach~\cite{Dupont2018,Horvatic2020}. We focus first on the isothermal $T_1^{-1} (B)$ dependence in the TLL regime. In the study of spin-ladder systems, it has been shown~\cite{Jeong2016} that at the temperature $T \approx 2T_c$ the $T_1 ^{-1}(B)$ dependence can reveal the nature of inherent fermionic interactions - a concave (convex) dependence directly reflects the attractive (repulsive) interactions in the system. The advantage of analysing $T_1 ^{-1}(B)$  is that it avoids possible ambiguities stemming from the delicate analysis of the $T_1 ^{-1}(T)$ data. It has long been theoretically predicted~\cite{Konik2002,Fath03} that for an $S=1$ Haldane chain system the quasiparticle interactions in the TLL phase are attractive, and thus we expect a concave field dependence. In Fig.~\ref{fig:fig3}(a) we show the $T_1 ^{-1}(B)$ measured at $6.5$~K ($>2 T_c^{max.} $) together with the previously reported data for the two ladder systems: (C$_7$H$_{10}$N)$_2$CuBr$_4$ (DIMPY)~\cite{Jeong2016} and (C$_5$H$_{12}$N)$_2$CuBr$_4$ (BPCB)~\cite{Klanjsek2008}, with attractive and repulsive interactions in the TLL regime, respectively. Clearly, BoNO displays attractive interactions of TLL quasiparticles. Apart from the qualitative behaviour, using the values of $K$, $u$, and $A_0 ^x$ calculated by DMRG (see Fig.~S8 in the SM~\cite{Supp}), we also quantitatively compare the experimental and numerical results, with $A_\perp$ as a single fitting parameter. As shown, the calculated curves follow the experimental data, albeit for $B$ close to \Bl\ and \Bu, where the system is close to the edges of the band,  the TLL description is not suitable. It can also be seen that at this temperature the RPA correction only has a minor effect (see Figs.~S9 and~S10 in the SM~\cite{Supp}), which confirms the validity of our approach. Our data clearly reveal the presence of attractive interactions in the TLL state of a Haldane spin chain system in a broad region of phase space. This is the first direct demonstration of the theoretically predicted $K(B)$ dependence, made more than two decades ago.\\
\indent We proceed to analyse our  $T_1 ^{-1}(T)$ data by applying the least-square (LS) fitting of the complete expression $T_1 ^{-1} = T_{1, \text{TLL}} ^{-1} \Phi(K, T_c/T)$, with $K$ and $A_\perp$ as fitting parameters, and $T_c$ determined from the experiment. Figs.~\ref{fig:fig3}(b) and (c) (and Fig.~S11 in the SM~\cite{Supp}) show an example of the fitted data at $19$~T and $27$~T. The resulting $K_\text{LS}$ values dependence across the phase diagram are shown as red circles in Fig.~\ref{fig:fig3}(d). Although the fitted $K$ values are consistent with a dome-shaped DMRG predictions, the large parameter uncertainties are the result of scarce data points due to the limited measurement time on the high-field magnet. Curiously, the fitted $T_1 ^{-1}(T)$ curves are almost indistinguishable from the ones calculated using theoretical $K$ values (Figs.~\ref{fig:fig3}(b) and (c)). Stability of LS fits depend on the number of measured points since the RPA correction function $\Phi(K, T_c/T)$ diverges at $T=T_c$, and the fitting procedure will dominantly minimize the square-error close to the divergence. We also treat our data using a BIA method (now with $T_c$ as a fitting parameter) which compensates the mentioned issue and is well suited for this case~\cite{VonToussaint}. In Figs.~\ref{fig:fig3}(b) and (c) we show the calculated probability distributions for the $T_1 ^{-1}(T)$ datasets, together with the mean values of the $T_1 ^{-1}(T)$ dependence and 95\%  HDI. In Figs.~\ref{fig:fig3}(d) and (e) we respectively show the resulting probability distribution for $K$ parameter and $T_c$, with the mean values and 68\% HDI. Our data show excellent agreement with theoretical predictions. This result serves as a consistency check of the $K(B)$ values obtained in the previous section and confirms the validity of the RPA-based correction~\cite{Dupont2018,Horvatic2020} to Haldane systems.
\\
\indent Finally, we focus on the $T_1 ^{-1} (B)$ data in the 1D regime for $B \approx \Bu$. With an increase of $B$ for $T < \jld$, the system crosses over from the gapless TLL into the gapped fully polarized phase through a QC region that expands in a characteristic $V$-shape~\cite{Maeda2007} above 0~K to higher temperatures. For $B=\Bu$, where the system is directly at the QCP, one expects to find~\cite{Mukhopadhyay2012} a universal and scale invariant  $T_1^{-1}$ dependence on $|B-B_\text{c1,c2}|/T$. Our data measured at 6.5~K are shown in Fig.~\ref{fig:fig3}(f) together with the data of the large-$D$ spin $S=1$ chain compound NiCl$_2$-4SC(NH$_2$)$_2$ (DTN) and the $S=1/2$ ladder system BPCB from Ref.~[\onlinecite{Mukhopadhyay2012}]. Data at 6.5~K were chosen to correspond to the relative energy scale of the 3D interaction ($T_\text{meas.}/T_{c}^{max.}$) of the latter systems. Measurements at other temperatures of BoNO are shown in Fig.~S13 in the SM~\cite{Supp}. Very close to \Bu, a steep $T_1 ^{-1} (B) \propto e^{-3\Delta(B)/T}$ dependence is observed, with $\Delta(B) = g \mu_B (B-\Bu)/k_B$. The latter is driven by the critical excitations, theoretically described by three-magnon propagators~\cite{Orignac2007} that have been shown to be very robust~\cite{Ranjith2022} -- perpetuating up to high temperatures. The scaled data for BoNO lie exactly on top of the other two systems, demonstrating the transition at \Bu\ is a true QCP. This perfect data overlap in three different systems shows that the universality is set by interaction-dependent scale factor, irrespective of the intrinsic nature of the system.\\
\indent In summary, we presented an extensive study of the magnetic-field induced gapless phases in the new organic Haldane $S=1$ chain $m$-NO$_2$PhBNO (abbreviated as BoNO) across the $B-T$ phase diagram. The system is completely described by the quasi-1D Heisenberg Hamiltonian, without additional spin-orbit-related terms, which means it can be used to study predictions of the pure Haldane chain system. Through the analysis of the TLL regime, we present the first confirmation of the theoretically predicted attractive interactions. At temperatures below 2.9~K, a long-range BEC-like order emerges, confirmed by the quantum Monte Carlo calculations, the critical exponent of $\nu = 0.66$ at \Bu\ and the quantum critical scaling that reveals a universally robust three-magnon spin-dynamics.

\subsection{End Matter}
\textit{Appendix A: Samples}--Single crystals of BoNO were grown using a synthesis process described in Ref.~\onlinecite{Jakovac2024}. Crystal structure and axes orientation was determined by a single-crystal X-ray diffraction using the Mo K$\alpha$ ($\lambda = 0.71$~\AA) radiation source. For experiments we used a sample with dimensions $ a \times b \times c = 2.7 \times 0.72 \times 0.1$~mm$^3$.\\
\\
\textit{Appendix B: Nuclear magnetic resonance}--$^{1}$H NMR measurements were performed using pulsed spectrometers. Data in magnetic fields up to 12~T were collected in an Oxford Instruments superconducting magnet in the NMR laboratory at University of Zagreb. Measurements at higher magnetic fields were performed in a 20~MW resistive magnet at high-magnetic field facility at LNCMI-CNRS in Grenoble. For the measurements up to 12~T the magnetic field was oriented parallel to the crystallographic $c$ axis. Due to the overlap of NMR lines it was difficult to trace the lines when entering the ordered phase. Therefore, for the measurements in the resistive magnet, the sample was oriented away from the crystallographic axes so that we can keep track of each $^1$H spectral line - the normalized magnetic field orientation was [0.373, 0.924, 0.077]. The spectra were collected by Hahn echo sequence $\pi/2 - \tau -\pi$ with a typical value of a $\pi/2$ pulse of 6 $\mu$s, and $\tau = 100~\mu$s. The $T_1$ measurements were performed using a saturation-recovery technique on a selected NMR line shown in Fig.~S3 in the SM~\cite{Supp}. Data were fitted to a magnetic relaxation function of the form: $M(t)=M_0 (1-e^{-(t/T_1)^{\beta}})$, where $\beta \approx 1$ in all measurements.\\
\\
\textit{Appendix C: Heat capacity}--Heat capacity was measured using a home-made setup with a resistive heater attached to the sample by proton-free vacuum grease from Daikin Industries. The heater–sample assembly was mounted in vacuum and thermally isolated from the rest of the apparatus by a thin support wire. Heating was performed by applying current pulses to the resistor, and the temperature of the heater was simultaneously monitored through its resistance. From the deposited energy and the corresponding temperature change, the total heat capacity of the heater–sample system was obtained. To isolate the sample contribution, reference measurements were performed under identical conditions without the sample. Subtracting these data yielded the sample's heat capacity. The measurements were not calibrated to absolute units; thus, the reported heat capacity is given in arbitrary units. Nevertheless, this method is well suited to detecting phase transitions, since the heat capacity must diverge at the transition temperature $T_c$ for a second-order phase transition. Therefore, the $T_c$ can be determined reliably even without an absolute calibration.\\
\\
\textit{Appendix D: Magnetostriction}--Magnetostriction measurements were performed in pulsed magnetic fields up to 41.2~T at the high-magnetic-field facility of HZDR in Dresden using the fiber Bragg grating (FBG) method~\cite{Daou2010}. Several samples with lengths of approximately $2$--$5$~mm and a cross-section of $0.5 \times 0.1$~mm$^2$ were used. The samples were glued to an $80~\mu\text{m}$ fiber using Loctite~401 cyanoacrylate, which has been shown to be safe for organic samples. The final data were averaged over three samples (a total of six measurements). The mechanical losses to the fiber were determined to be negligible, $(Y_{\text{fiber}} A_{\text{fiber}}) / (Y_{\text{sample}} A_{\text{sample}}) < 3\%$.\\
\\
\textit{Appendix E: Data analysis}--Quantitative analysis was carried out using a combination of least-squares (LS) fitting and Bayesian inference analysis (BIA). LS fitting minimizes the error function ($\chi^2$) or maximizes the likelihood $p(x|\theta)$ of observing data $x$ with the parameter vector $\theta$ in our hypothetical physical phenomena $f(x,\theta)$, yielding a single parameter vector $\theta$ as a point estimate.  While straightforward, this approach is sensitive to noise and often converges only to local optima. BIA, by contrast, applies Bayes’ theorem $p(\theta|x) \propto p(x|\theta)p(\theta)$ to generate a posterior probability distributions for the parameters. The question is now inverted - $p(\theta|x)$ tells us how likely we have parameters $\theta$, given that we have just observed data $x$. This method provides both point estimates and confidence intervals, and is generally more robust when the dataset is limited or noisy, though results depend on the choice of prior distribution $p(\theta)$.\\
\indent To facilitate numerical convergence in fitting $T_1^{-1}$, we first tabulated the multiplicative correction factor $\Phi(K, T_c/T)$ to the required precision~\cite{Horvatic2020}. Bayesian sampling was performed with the PyMC5 library. The priors for $K$ and $T_c$ were set as truncated normal distributions centered at $K_\text{DMRG}$ and $T_c$ extracted from $T_1^{-1}$, respectively. The widths were chosen broad enough to avoid biasing the fits, with truncation ranges $[0,3]$ for $K$ and $[0,10]$~K for $T_c$. The posterior distributions were constructed by drawing 5000 samples from 8 independent Markov chains.  LS fitting used the same starting values but failed to converge at certain magnetic fields, in contrast to the Bayesian approach, which yielded stable posterior distributions and highest density ranges for all parameters.\\
\\
\textit{Appendix F: DMRG and QMC data}--For one dimension, the TLL properties were computed using the density-matrix renormalization group (DMRG) method within the matrix product state formalism~\cite{Schollwock} based on ITensor library~\cite{itensor}, analoguous to the one performed on DTN material~\cite{Blinder2017}. Finite-temperature and thermodynamic properties of the three dimensional problem were obtained from quantum Monte Carlo simulations based on the stochastic series expansion algorithm with directed-loop updates~\cite{Syljuasen}. Large system sizes were simulated to access the thermodynamic limit, and statistical uncertainties were estimated from binning analysis after sufficient thermalization. A standard finite-size scaling analysis of the order parameter~\cite{Sandvik} was employed to extract the bulk 3D critical temperatures across the entire field range of the phase diagram shown in Fig.~\ref{fig:fig2}(d).\\
\\


\begin{thebibliography}{79}%
	\makeatletter
	\providecommand \@ifxundefined [1]{%
		\@ifx{#1\undefined}
	}%
	\providecommand \@ifnum [1]{%
		\ifnum #1\expandafter \@firstoftwo
		\else \expandafter \@secondoftwo
		\fi
	}%
	\providecommand \@ifx [1]{%
		\ifx #1\expandafter \@firstoftwo
		\else \expandafter \@secondoftwo
		\fi
	}%
	\providecommand \natexlab [1]{#1}%
	\providecommand \enquote  [1]{``#1''}%
	\providecommand \bibnamefont  [1]{#1}%
	\providecommand \bibfnamefont [1]{#1}%
	\providecommand \citenamefont [1]{#1}%
	\providecommand \href@noop [0]{\@secondoftwo}%
	\providecommand \href [0]{\begingroup \@sanitize@url \@href}%
	\providecommand \@href[1]{\@@startlink{#1}\@@href}%
	\providecommand \@@href[1]{\endgroup#1\@@endlink}%
	\providecommand \@sanitize@url [0]{\catcode `\\12\catcode `\$12\catcode
		`\&12\catcode `\#12\catcode `\^12\catcode `\_12\catcode `\%12\relax}%
	\providecommand \@@startlink[1]{}%
	\providecommand \@@endlink[0]{}%
	\providecommand \url  [0]{\begingroup\@sanitize@url \@url }%
	\providecommand \@url [1]{\endgroup\@href {#1}{\urlprefix }}%
	\providecommand \urlprefix  [0]{URL }%
	\providecommand \Eprint [0]{\href }%
	\providecommand \doibase [0]{http://dx.doi.org/}%
	\providecommand \selectlanguage [0]{\@gobble}%
	\providecommand \bibinfo  [0]{\@secondoftwo}%
	\providecommand \bibfield  [0]{\@secondoftwo}%
	\providecommand \translation [1]{[#1]}%
	\providecommand \BibitemOpen [0]{}%
	\providecommand \bibitemStop [0]{}%
	\providecommand \bibitemNoStop [0]{.\EOS\space}%
	\providecommand \EOS [0]{\spacefactor3000\relax}%
	\providecommand \BibitemShut  [1]{\csname bibitem#1\endcsname}%
	\let\auto@bib@innerbib\@empty
	\bibitem [{\citenamefont {Giamarchi}(2003)}]{GiamarchiBook}%
	\BibitemOpen
	\bibfield  {author} {\bibinfo {author} {\bibfnamefont {T.}~\bibnamefont
			{Giamarchi}},\ }\href {\doibase 10.1093/acprof:oso/9780198525004.001.0001}
	{\emph {\bibinfo {title} {{Quantum Physics in One Dimension}}}}\ (\bibinfo
	{publisher} {Oxford University Press},\ \bibinfo {year} {2003})\BibitemShut
	{NoStop}%
	\bibitem [{\citenamefont {Zapf}\ \emph {et~al.}(2014)\citenamefont {Zapf},
		\citenamefont {Jaime},\ and\ \citenamefont {Batista}}]{Zapf2014}%
	\BibitemOpen
	\bibfield  {author} {\bibinfo {author} {\bibfnamefont {V.}~\bibnamefont
			{Zapf}}, \bibinfo {author} {\bibfnamefont {M.}~\bibnamefont {Jaime}}, \ and\
		\bibinfo {author} {\bibfnamefont {C.~D.}\ \bibnamefont {Batista}},\ }\href
	{\doibase 10.1103/RevModPhys.86.563} {\bibfield  {journal} {\bibinfo
			{journal} {Reviews of Modern Physics}\ }\textbf {\bibinfo {volume} {86}},\
		\bibinfo {pages} {563} (\bibinfo {year} {2014})}\BibitemShut {NoStop}%
	\bibitem [{\citenamefont {Klanjšek}\ \emph {et~al.}(2008)\citenamefont
		{Klanjšek}, \citenamefont {Mayaffre}, \citenamefont {Berthier},
		\citenamefont {Horvatić}, \citenamefont {Chiari}, \citenamefont {Piovesana},
		\citenamefont {Bouillot}, \citenamefont {Kollath}, \citenamefont {Orignac},
		\citenamefont {Citro},\ and\ \citenamefont {Giamarchi}}]{Klanjsek2008}%
	\BibitemOpen
	\bibfield  {author} {\bibinfo {author} {\bibfnamefont {M.}~\bibnamefont
			{Klanjšek}}, \bibinfo {author} {\bibfnamefont {H.}~\bibnamefont {Mayaffre}},
		\bibinfo {author} {\bibfnamefont {C.}~\bibnamefont {Berthier}}, \bibinfo
		{author} {\bibfnamefont {M.}~\bibnamefont {Horvatić}}, \bibinfo {author}
		{\bibfnamefont {B.}~\bibnamefont {Chiari}}, \bibinfo {author} {\bibfnamefont
			{O.}~\bibnamefont {Piovesana}}, \bibinfo {author} {\bibfnamefont
			{P.}~\bibnamefont {Bouillot}}, \bibinfo {author} {\bibfnamefont
			{C.}~\bibnamefont {Kollath}}, \bibinfo {author} {\bibfnamefont
			{E.}~\bibnamefont {Orignac}}, \bibinfo {author} {\bibfnamefont
			{R.}~\bibnamefont {Citro}}, \ and\ \bibinfo {author} {\bibfnamefont
			{T.}~\bibnamefont {Giamarchi}},\ }\href {\doibase
		10.1103/PhysRevLett.101.137207} {\bibfield  {journal} {\bibinfo  {journal}
			{Physical Review Letters}\ }\textbf {\bibinfo {volume} {101}},\ \bibinfo
		{pages} {137207} (\bibinfo {year} {2008})}\BibitemShut {NoStop}%
	\bibitem [{\citenamefont {Jeong}\ \emph {et~al.}(2013)\citenamefont {Jeong},
		\citenamefont {Mayaffre}, \citenamefont {Berthier}, \citenamefont
		{Schmidiger}, \citenamefont {Zheludev},\ and\ \citenamefont
		{Horvati\ifmmode~\acute{c}\else \'{c}\fi{}}}]{Jeong2013}%
	\BibitemOpen
	\bibfield  {author} {\bibinfo {author} {\bibfnamefont {M.}~\bibnamefont
			{Jeong}}, \bibinfo {author} {\bibfnamefont {H.}~\bibnamefont {Mayaffre}},
		\bibinfo {author} {\bibfnamefont {C.}~\bibnamefont {Berthier}}, \bibinfo
		{author} {\bibfnamefont {D.}~\bibnamefont {Schmidiger}}, \bibinfo {author}
		{\bibfnamefont {A.}~\bibnamefont {Zheludev}}, \ and\ \bibinfo {author}
		{\bibfnamefont {M.}~\bibnamefont {Horvati\ifmmode~\acute{c}\else
				\'{c}\fi{}}},\ }\href {\doibase 10.1103/PhysRevLett.111.106404} {\bibfield
		{journal} {\bibinfo  {journal} {Physical Review Letters}\ }\textbf {\bibinfo
			{volume} {111}},\ \bibinfo {pages} {106404} (\bibinfo {year}
		{2013})}\BibitemShut {NoStop}%
	\bibitem [{\citenamefont {Vijayan}\ \emph {et~al.}(2020)\citenamefont
		{Vijayan}, \citenamefont {Sompet}, \citenamefont {Salomon}, \citenamefont
		{Koepsell}, \citenamefont {Hirthe}, \citenamefont {Bohrdt}, \citenamefont
		{Grusdt}, \citenamefont {Bloch},\ and\ \citenamefont {Gross}}]{Vijayan}%
	\BibitemOpen
	\bibfield  {author} {\bibinfo {author} {\bibfnamefont {J.}~\bibnamefont
			{Vijayan}}, \bibinfo {author} {\bibfnamefont {P.}~\bibnamefont {Sompet}},
		\bibinfo {author} {\bibfnamefont {G.}~\bibnamefont {Salomon}}, \bibinfo
		{author} {\bibfnamefont {J.}~\bibnamefont {Koepsell}}, \bibinfo {author}
		{\bibfnamefont {S.}~\bibnamefont {Hirthe}}, \bibinfo {author} {\bibfnamefont
			{A.}~\bibnamefont {Bohrdt}}, \bibinfo {author} {\bibfnamefont
			{F.}~\bibnamefont {Grusdt}}, \bibinfo {author} {\bibfnamefont
			{I.}~\bibnamefont {Bloch}}, \ and\ \bibinfo {author} {\bibfnamefont
			{C.}~\bibnamefont {Gross}},\ }\href {\doibase 10.1126/science.aay2354}
	{\bibfield  {journal} {\bibinfo  {journal} {Science}\ }\textbf {\bibinfo
			{volume} {367}},\ \bibinfo {pages} {186} (\bibinfo {year}
		{2020})}\BibitemShut {NoStop}%
	\bibitem [{\citenamefont {Wu}\ \emph {et~al.}(2019)\citenamefont {Wu},
		\citenamefont {Nikitin}, \citenamefont {Wang}, \citenamefont {Zhu},
		\citenamefont {Batista}, \citenamefont {Tsvelik}, \citenamefont {Samarakoon},
		\citenamefont {Tennant}, \citenamefont {Brando}, \citenamefont {Vasylechko},
		\citenamefont {Frontzek}, \citenamefont {Savici}, \citenamefont {Sala},
		\citenamefont {Ehlers}, \citenamefont {Christianson}, \citenamefont
		{Lumsden},\ and\ \citenamefont {Podlesnyak}}]{Wu2019}%
	\BibitemOpen
	\bibfield  {author} {\bibinfo {author} {\bibfnamefont {L.~S.}\ \bibnamefont
			{Wu}}, \bibinfo {author} {\bibfnamefont {S.~E.}\ \bibnamefont {Nikitin}},
		\bibinfo {author} {\bibfnamefont {Z.}~\bibnamefont {Wang}}, \bibinfo {author}
		{\bibfnamefont {W.}~\bibnamefont {Zhu}}, \bibinfo {author} {\bibfnamefont
			{C.~D.}\ \bibnamefont {Batista}}, \bibinfo {author} {\bibfnamefont {A.~M.}\
			\bibnamefont {Tsvelik}}, \bibinfo {author} {\bibfnamefont {A.~M.}\
			\bibnamefont {Samarakoon}}, \bibinfo {author} {\bibfnamefont {D.~A.}\
			\bibnamefont {Tennant}}, \bibinfo {author} {\bibfnamefont {M.}~\bibnamefont
			{Brando}}, \bibinfo {author} {\bibfnamefont {L.}~\bibnamefont {Vasylechko}},
		\bibinfo {author} {\bibfnamefont {M.}~\bibnamefont {Frontzek}}, \bibinfo
		{author} {\bibfnamefont {A.~T.}\ \bibnamefont {Savici}}, \bibinfo {author}
		{\bibfnamefont {G.}~\bibnamefont {Sala}}, \bibinfo {author} {\bibfnamefont
			{G.}~\bibnamefont {Ehlers}}, \bibinfo {author} {\bibfnamefont {A.~D.}\
			\bibnamefont {Christianson}}, \bibinfo {author} {\bibfnamefont {M.~D.}\
			\bibnamefont {Lumsden}}, \ and\ \bibinfo {author} {\bibfnamefont
			{A.}~\bibnamefont {Podlesnyak}},\ }\href {\doibase
		10.1038/s41467-019-08485-7} {\bibfield  {journal} {\bibinfo  {journal}
			{Nature Communications}\ }\textbf {\bibinfo {volume} {10}},\ \bibinfo {pages}
		{698} (\bibinfo {year} {2019})}\BibitemShut {NoStop}%
	\bibitem [{\citenamefont {Bouchoule}\ \emph {et~al.}(2025)\citenamefont
		{Bouchoule}, \citenamefont {Citro}, \citenamefont {Duty}, \citenamefont
		{Giamarchi}, \citenamefont {Hulet}, \citenamefont {Klanj{\v{s}}ek},
		\citenamefont {Orignac},\ and\ \citenamefont {Weber}}]{Bouchoule2025}%
	\BibitemOpen
	\bibfield  {author} {\bibinfo {author} {\bibfnamefont {I.}~\bibnamefont
			{Bouchoule}}, \bibinfo {author} {\bibfnamefont {R.}~\bibnamefont {Citro}},
		\bibinfo {author} {\bibfnamefont {T.}~\bibnamefont {Duty}}, \bibinfo {author}
		{\bibfnamefont {T.}~\bibnamefont {Giamarchi}}, \bibinfo {author}
		{\bibfnamefont {R.~G.}\ \bibnamefont {Hulet}}, \bibinfo {author}
		{\bibfnamefont {M.}~\bibnamefont {Klanj{\v{s}}ek}}, \bibinfo {author}
		{\bibfnamefont {E.}~\bibnamefont {Orignac}}, \ and\ \bibinfo {author}
		{\bibfnamefont {B.}~\bibnamefont {Weber}},\ }\href {\doibase
		10.1038/s42254-025-00866-w} {\bibfield  {journal} {\bibinfo  {journal}
			{Nature Reviews Physics}\ }\textbf {\bibinfo {volume} {7}},\ \bibinfo {pages}
		{565} (\bibinfo {year} {2025})}\BibitemShut {NoStop}%
	\bibitem [{\citenamefont {Keimer}\ and\ \citenamefont
		{Moore}(2017)}]{Keimer2017}%
	\BibitemOpen
	\bibfield  {author} {\bibinfo {author} {\bibfnamefont {B.}~\bibnamefont
			{Keimer}}\ and\ \bibinfo {author} {\bibfnamefont {J.~E.}\ \bibnamefont
			{Moore}},\ }\href {\doibase 10.1038/nphys4302} {\bibfield  {journal}
		{\bibinfo  {journal} {Nature Physics}\ }\textbf {\bibinfo {volume} {13}},\
		\bibinfo {pages} {1045} (\bibinfo {year} {2017})}\BibitemShut {NoStop}%
	\bibitem [{\citenamefont {Autti}\ \emph {et~al.}(2018)\citenamefont {Autti},
		\citenamefont {Eltsov},\ and\ \citenamefont {Volovik}}]{Autti}%
	\BibitemOpen
	\bibfield  {author} {\bibinfo {author} {\bibfnamefont {S.}~\bibnamefont
			{Autti}}, \bibinfo {author} {\bibfnamefont {V.~B.}\ \bibnamefont {Eltsov}}, \
		and\ \bibinfo {author} {\bibfnamefont {G.~E.}\ \bibnamefont {Volovik}},\
	}\href {\doibase 10.1103/PhysRevLett.120.215301} {\bibfield  {journal}
		{\bibinfo  {journal} {Physical Review Letters}\ }\textbf {\bibinfo {volume}
			{120}},\ \bibinfo {pages} {215301} (\bibinfo {year} {2018})}\BibitemShut
	{NoStop}%
	\bibitem [{\citenamefont {Aveline}\ \emph {et~al.}(2020)\citenamefont
		{Aveline}, \citenamefont {Williams}, \citenamefont {Elliott}, \citenamefont
		{Dutenhoffer}, \citenamefont {Kellogg}, \citenamefont {Kohel}, \citenamefont
		{Lay}, \citenamefont {Oudrhiri}, \citenamefont {Shotwell}, \citenamefont
		{Yu},\ and\ \citenamefont {Thompson}}]{Aveline}%
	\BibitemOpen
	\bibfield  {author} {\bibinfo {author} {\bibfnamefont {D.~C.}\ \bibnamefont
			{Aveline}}, \bibinfo {author} {\bibfnamefont {J.~R.}\ \bibnamefont
			{Williams}}, \bibinfo {author} {\bibfnamefont {E.~R.}\ \bibnamefont
			{Elliott}}, \bibinfo {author} {\bibfnamefont {C.}~\bibnamefont
			{Dutenhoffer}}, \bibinfo {author} {\bibfnamefont {J.~R.}\ \bibnamefont
			{Kellogg}}, \bibinfo {author} {\bibfnamefont {J.~M.}\ \bibnamefont {Kohel}},
		\bibinfo {author} {\bibfnamefont {N.~E.}\ \bibnamefont {Lay}}, \bibinfo
		{author} {\bibfnamefont {K.}~\bibnamefont {Oudrhiri}}, \bibinfo {author}
		{\bibfnamefont {R.~F.}\ \bibnamefont {Shotwell}}, \bibinfo {author}
		{\bibfnamefont {N.}~\bibnamefont {Yu}}, \ and\ \bibinfo {author}
		{\bibfnamefont {R.~J.}\ \bibnamefont {Thompson}},\ }\href {\doibase
		10.1038/s41586-020-2346-1} {\bibfield  {journal} {\bibinfo  {journal}
			{Nature}\ }\textbf {\bibinfo {volume} {582}},\ \bibinfo {pages} {193}
		(\bibinfo {year} {2020})}\BibitemShut {NoStop}%
	\bibitem [{\citenamefont {Sompet}\ \emph {et~al.}(2022)\citenamefont {Sompet},
		\citenamefont {Hirthe}, \citenamefont {Bourgund}, \citenamefont {Chalopin},
		\citenamefont {Bibo}, \citenamefont {Koepsell}, \citenamefont
		{Bojovi{\'{c}}}, \citenamefont {Verresen}, \citenamefont {Pollmann},
		\citenamefont {Salomon}, \citenamefont {Gross}, \citenamefont {Hilker},\ and\
		\citenamefont {Bloch}}]{Sompet2022}%
	\BibitemOpen
	\bibfield  {author} {\bibinfo {author} {\bibfnamefont {P.}~\bibnamefont
			{Sompet}}, \bibinfo {author} {\bibfnamefont {S.}~\bibnamefont {Hirthe}},
		\bibinfo {author} {\bibfnamefont {D.}~\bibnamefont {Bourgund}}, \bibinfo
		{author} {\bibfnamefont {T.}~\bibnamefont {Chalopin}}, \bibinfo {author}
		{\bibfnamefont {J.}~\bibnamefont {Bibo}}, \bibinfo {author} {\bibfnamefont
			{J.}~\bibnamefont {Koepsell}}, \bibinfo {author} {\bibfnamefont
			{P.}~\bibnamefont {Bojovi{\'{c}}}}, \bibinfo {author} {\bibfnamefont
			{R.}~\bibnamefont {Verresen}}, \bibinfo {author} {\bibfnamefont
			{F.}~\bibnamefont {Pollmann}}, \bibinfo {author} {\bibfnamefont
			{G.}~\bibnamefont {Salomon}}, \bibinfo {author} {\bibfnamefont
			{C.}~\bibnamefont {Gross}}, \bibinfo {author} {\bibfnamefont {T.~A.}\
			\bibnamefont {Hilker}}, \ and\ \bibinfo {author} {\bibfnamefont
			{I.}~\bibnamefont {Bloch}},\ }\href {\doibase 10.1038/s41586-022-04688-z}
	{\bibfield  {journal} {\bibinfo  {journal} {Nature}\ }\textbf {\bibinfo
			{volume} {606}},\ \bibinfo {pages} {484} (\bibinfo {year}
		{2022})}\BibitemShut {NoStop}%
	\bibitem [{\citenamefont {Byrnes}\ \emph {et~al.}(2012)\citenamefont {Byrnes},
		\citenamefont {Wen},\ and\ \citenamefont {Yamamoto}}]{Byrnes}%
	\BibitemOpen
	\bibfield  {author} {\bibinfo {author} {\bibfnamefont {T.}~\bibnamefont
			{Byrnes}}, \bibinfo {author} {\bibfnamefont {K.}~\bibnamefont {Wen}}, \ and\
		\bibinfo {author} {\bibfnamefont {Y.}~\bibnamefont {Yamamoto}},\ }\href
	{\doibase 10.1103/PhysRevA.85.040306} {\bibfield  {journal} {\bibinfo
			{journal} {Physical Review A}\ }\textbf {\bibinfo {volume} {85}},\ \bibinfo
		{pages} {040306} (\bibinfo {year} {2012})}\BibitemShut {NoStop}%
	\bibitem [{\citenamefont {Andrianov}\ and\ \citenamefont
		{Moiseev}(2014)}]{Andrianov}%
	\BibitemOpen
	\bibfield  {author} {\bibinfo {author} {\bibfnamefont {S.~N.}\ \bibnamefont
			{Andrianov}}\ and\ \bibinfo {author} {\bibfnamefont {S.~A.}\ \bibnamefont
			{Moiseev}},\ }\href {\doibase 10.1103/PhysRevA.90.042303} {\bibfield
		{journal} {\bibinfo  {journal} {Physical Review A}\ }\textbf {\bibinfo
			{volume} {90}},\ \bibinfo {pages} {042303} (\bibinfo {year}
		{2014})}\BibitemShut {NoStop}%
	\bibitem [{\citenamefont {Mohseni}\ \emph {et~al.}(2022)\citenamefont
		{Mohseni}, \citenamefont {Vasyuchka}, \citenamefont {L'vov}, \citenamefont
		{Serga},\ and\ \citenamefont {Hillebrands}}]{Mohseni2022}%
	\BibitemOpen
	\bibfield  {author} {\bibinfo {author} {\bibfnamefont {M.}~\bibnamefont
			{Mohseni}}, \bibinfo {author} {\bibfnamefont {V.~I.}\ \bibnamefont
			{Vasyuchka}}, \bibinfo {author} {\bibfnamefont {V.~S.}\ \bibnamefont
			{L'vov}}, \bibinfo {author} {\bibfnamefont {A.~A.}\ \bibnamefont {Serga}}, \
		and\ \bibinfo {author} {\bibfnamefont {B.}~\bibnamefont {Hillebrands}},\
	}\href {\doibase 10.1038/s42005-022-00970-8} {\bibfield  {journal} {\bibinfo
			{journal} {Communications Physics}\ }\textbf {\bibinfo {volume} {5}},\
		\bibinfo {pages} {196} (\bibinfo {year} {2022})}\BibitemShut {NoStop}%
	\bibitem [{\citenamefont {Haldane}(1983)}]{HaldanePRL}%
	\BibitemOpen
	\bibfield  {author} {\bibinfo {author} {\bibfnamefont {F.~D.~M.}\
			\bibnamefont {Haldane}},\ }\href {\doibase 10.1103/PhysRevLett.50.1153}
	{\bibfield  {journal} {\bibinfo  {journal} {Physical Review Letters}\
		}\textbf {\bibinfo {volume} {50}},\ \bibinfo {pages} {1153} (\bibinfo {year}
		{1983})}\BibitemShut {NoStop}%
	\bibitem [{\citenamefont {Sakai}\ and\ \citenamefont
		{Takahashi}(1990)}]{Sakai1990}%
	\BibitemOpen
	\bibfield  {author} {\bibinfo {author} {\bibfnamefont {T.}~\bibnamefont
			{Sakai}}\ and\ \bibinfo {author} {\bibfnamefont {M.}~\bibnamefont
			{Takahashi}},\ }\href {\doibase 10.1103/PhysRevB.42.4537} {\bibfield
		{journal} {\bibinfo  {journal} {Physical Review B}\ }\textbf {\bibinfo
			{volume} {42}},\ \bibinfo {pages} {4537} (\bibinfo {year}
		{1990})}\BibitemShut {NoStop}%
	\bibitem [{\citenamefont {Wierschem}\ and\ \citenamefont
		{Sengupta}(2014{\natexlab{a}})}]{WierschemPRL}%
	\BibitemOpen
	\bibfield  {author} {\bibinfo {author} {\bibfnamefont {K.}~\bibnamefont
			{Wierschem}}\ and\ \bibinfo {author} {\bibfnamefont {P.}~\bibnamefont
			{Sengupta}},\ }\href {\doibase 10.1103/PhysRevLett.112.247203} {\bibfield
		{journal} {\bibinfo  {journal} {Physical Review Letters}\ }\textbf {\bibinfo
			{volume} {112}},\ \bibinfo {pages} {247203} (\bibinfo {year}
		{2014}{\natexlab{a}})}\BibitemShut {NoStop}%
	\bibitem [{\citenamefont {Maximova}\ \emph {et~al.}(2021)\citenamefont
		{Maximova}, \citenamefont {Streltsov},\ and\ \citenamefont
		{Vasiliev}}]{Maximova}%
	\BibitemOpen
	\bibfield  {author} {\bibinfo {author} {\bibfnamefont {O.~V.}\ \bibnamefont
			{Maximova}}, \bibinfo {author} {\bibfnamefont {S.~V.}\ \bibnamefont
			{Streltsov}}, \ and\ \bibinfo {author} {\bibfnamefont {A.~N.}\ \bibnamefont
			{Vasiliev}},\ }\href {\doibase 10.1080/10408436.2020.1852911} {\bibfield
		{journal} {\bibinfo  {journal} {Critical Reviews in Solid State and Materials
				Sciences}\ }\textbf {\bibinfo {volume} {46}},\ \bibinfo {pages} {371}
		(\bibinfo {year} {2021})}\BibitemShut {NoStop}%
	\bibitem [{Sup()}]{Supp}%
	\BibitemOpen
	\href@noop {} {}\bibinfo {note} {See {Supplemental Material} at [URL will be
		inserted by publisher] for the list of previously studied {Haldane} systems,
		structure of the {BoNO} molecule, {$^1$}H {NMR} spectra, splitting of the
		{NMR} spectrum in the ordered phase, determination of the upper critical
		field {$B_{c2}$} using a windowing method, discussion on the magnetisation
		dependent intrachain/interchain coupling, {DMRG}-calculated values of {TLL}
		parameters, additional {$1/T_1(T)$} fits and relaxation curves, which
		includes {Refs}.~[20–49].}\BibitemShut {Stop}%
	\bibitem [{\citenamefont {Broholm}\ \emph {et~al.}(2001)\citenamefont
		{Broholm}, \citenamefont {Aeppli}, \citenamefont {Chen}, \citenamefont
		{Dender}, \citenamefont {Enderle}, \citenamefont {Hammar}, \citenamefont
		{Honda}, \citenamefont {Katsumata}, \citenamefont {Landee}, \citenamefont
		{Oshikawa}, \citenamefont {Regnault}, \citenamefont {Reich}, \citenamefont
		{Shapiro}, \citenamefont {Sieling}, \citenamefont {Stone}, \citenamefont
		{Turnbull}, \citenamefont {Zaliznyak},\ and\ \citenamefont
		{Zheludev}}]{Broholm2002}%
	\BibitemOpen
	\bibfield  {author} {\bibinfo {author} {\bibfnamefont {C.}~\bibnamefont
			{Broholm}}, \bibinfo {author} {\bibfnamefont {G.}~\bibnamefont {Aeppli}},
		\bibinfo {author} {\bibfnamefont {Y.}~\bibnamefont {Chen}}, \bibinfo {author}
		{\bibfnamefont {D.~C.}\ \bibnamefont {Dender}}, \bibinfo {author}
		{\bibfnamefont {M.}~\bibnamefont {Enderle}}, \bibinfo {author} {\bibfnamefont
			{P.~R.}\ \bibnamefont {Hammar}}, \bibinfo {author} {\bibfnamefont
			{Z.}~\bibnamefont {Honda}}, \bibinfo {author} {\bibfnamefont
			{K.}~\bibnamefont {Katsumata}}, \bibinfo {author} {\bibfnamefont {C.~P.}\
			\bibnamefont {Landee}}, \bibinfo {author} {\bibfnamefont {M.}~\bibnamefont
			{Oshikawa}}, \bibinfo {author} {\bibfnamefont {L.~P.}\ \bibnamefont
			{Regnault}}, \bibinfo {author} {\bibfnamefont {D.~H.}\ \bibnamefont {Reich}},
		\bibinfo {author} {\bibfnamefont {S.~M.}\ \bibnamefont {Shapiro}}, \bibinfo
		{author} {\bibfnamefont {M.}~\bibnamefont {Sieling}}, \bibinfo {author}
		{\bibfnamefont {M.~B.}\ \bibnamefont {Stone}}, \bibinfo {author}
		{\bibfnamefont {M.~M.}\ \bibnamefont {Turnbull}}, \bibinfo {author}
		{\bibfnamefont {I.}~\bibnamefont {Zaliznyak}}, \ and\ \bibinfo {author}
		{\bibfnamefont {A.}~\bibnamefont {Zheludev}},\ }in\ \href {\doibase
		10.1007/3-540-45649-X_8} {\emph {\bibinfo {booktitle} {"High Magnetic Fields:
				Applications in Condensed Matter Physics and Spectroscopy"}}},\ \bibinfo
	{editor} {edited by\ \bibinfo {editor} {\bibfnamefont {C.}~\bibnamefont
			{Berthier}}, \bibinfo {editor} {\bibfnamefont {L.~P.}\ \bibnamefont
			{L{\'e}vy}}, \ and\ \bibinfo {editor} {\bibfnamefont {G.}~\bibnamefont
			{Martinez}}}\ (\bibinfo  {publisher} {Springer Berlin Heidelberg},\ \bibinfo
	{address} {Berlin, Heidelberg},\ \bibinfo {year} {2001})\ Chap.~\bibinfo
	{chapter} {8}, pp.\ \bibinfo {pages} {211--234}\BibitemShut {NoStop}%
	\bibitem [{\citenamefont {Darriet}\ and\ \citenamefont
		{Regnault}(1993)}]{DARRIET1993}%
	\BibitemOpen
	\bibfield  {author} {\bibinfo {author} {\bibfnamefont {J.}~\bibnamefont
			{Darriet}}\ and\ \bibinfo {author} {\bibfnamefont {L.}~\bibnamefont
			{Regnault}},\ }\href {\doibase https://doi.org/10.1016/0038-1098(93)90455-V}
	{\bibfield  {journal} {\bibinfo  {journal} {Solid State Communications}\
		}\textbf {\bibinfo {volume} {86}},\ \bibinfo {pages} {409} (\bibinfo {year}
		{1993})}\BibitemShut {NoStop}%
	\bibitem [{\citenamefont {Xu}\ \emph {et~al.}(1996)\citenamefont {Xu},
		\citenamefont {DiTusa}, \citenamefont {Ito}, \citenamefont {Oka},
		\citenamefont {Takagi}, \citenamefont {Broholm},\ and\ \citenamefont
		{Aeppli}}]{Xu}%
	\BibitemOpen
	\bibfield  {author} {\bibinfo {author} {\bibfnamefont {G.}~\bibnamefont
			{Xu}}, \bibinfo {author} {\bibfnamefont {J.~F.}\ \bibnamefont {DiTusa}},
		\bibinfo {author} {\bibfnamefont {T.}~\bibnamefont {Ito}}, \bibinfo {author}
		{\bibfnamefont {K.}~\bibnamefont {Oka}}, \bibinfo {author} {\bibfnamefont
			{H.}~\bibnamefont {Takagi}}, \bibinfo {author} {\bibfnamefont
			{C.}~\bibnamefont {Broholm}}, \ and\ \bibinfo {author} {\bibfnamefont
			{G.}~\bibnamefont {Aeppli}},\ }\href {\doibase 10.1103/PhysRevB.54.R6827}
	{\bibfield  {journal} {\bibinfo  {journal} {Physical Review B}\ }\textbf
		{\bibinfo {volume} {54}},\ \bibinfo {pages} {R6827} (\bibinfo {year}
		{1996})}\BibitemShut {NoStop}%
	\bibitem [{\citenamefont {Takeuchi}\ \emph
		{et~al.}(1992{\natexlab{a}})\citenamefont {Takeuchi}, \citenamefont {Hori},
		\citenamefont {Yosida}, \citenamefont {Yamagishi}, \citenamefont {Katsumata},
		\citenamefont {Renard}, \citenamefont {Gadet}, \citenamefont {Verdaguer},\
		and\ \citenamefont {Date}}]{Takeuchi92}%
	\BibitemOpen
	\bibfield  {author} {\bibinfo {author} {\bibfnamefont {T.}~\bibnamefont
			{Takeuchi}}, \bibinfo {author} {\bibfnamefont {H.}~\bibnamefont {Hori}},
		\bibinfo {author} {\bibfnamefont {T.}~\bibnamefont {Yosida}}, \bibinfo
		{author} {\bibfnamefont {A.}~\bibnamefont {Yamagishi}}, \bibinfo {author}
		{\bibfnamefont {K.}~\bibnamefont {Katsumata}}, \bibinfo {author}
		{\bibfnamefont {J.-P.}\ \bibnamefont {Renard}}, \bibinfo {author}
		{\bibfnamefont {V.}~\bibnamefont {Gadet}}, \bibinfo {author} {\bibfnamefont
			{M.}~\bibnamefont {Verdaguer}}, \ and\ \bibinfo {author} {\bibfnamefont
			{M.}~\bibnamefont {Date}},\ }\href {\doibase 10.1143/JPSJ.61.3262} {\bibfield
		{journal} {\bibinfo  {journal} {Journal of the Physical Society of Japan}\
		}\textbf {\bibinfo {volume} {61}},\ \bibinfo {pages} {3262} (\bibinfo {year}
		{1992}{\natexlab{a}})}\BibitemShut {NoStop}%
	\bibitem [{\citenamefont {Zheludev}\ \emph {et~al.}(1996)\citenamefont
		{Zheludev}, \citenamefont {Nagler}, \citenamefont {Shapiro}, \citenamefont
		{Chou}, \citenamefont {Talham},\ and\ \citenamefont {Meisel}}]{Zheludev96}%
	\BibitemOpen
	\bibfield  {author} {\bibinfo {author} {\bibfnamefont {A.}~\bibnamefont
			{Zheludev}}, \bibinfo {author} {\bibfnamefont {S.~E.}\ \bibnamefont
			{Nagler}}, \bibinfo {author} {\bibfnamefont {S.~M.}\ \bibnamefont {Shapiro}},
		\bibinfo {author} {\bibfnamefont {L.~K.}\ \bibnamefont {Chou}}, \bibinfo
		{author} {\bibfnamefont {D.~R.}\ \bibnamefont {Talham}}, \ and\ \bibinfo
		{author} {\bibfnamefont {M.~W.}\ \bibnamefont {Meisel}},\ }\href {\doibase
		10.1103/PhysRevB.53.15004} {\bibfield  {journal} {\bibinfo  {journal}
			{Physical Review B}\ }\textbf {\bibinfo {volume} {53}},\ \bibinfo {pages}
		{15004} (\bibinfo {year} {1996})}\BibitemShut {NoStop}%
	\bibitem [{\citenamefont {Takeuchi}\ \emph
		{et~al.}(1992{\natexlab{b}})\citenamefont {Takeuchi}, \citenamefont {Ono},
		\citenamefont {Hori}, \citenamefont {Yosida}, \citenamefont {Yamagishi},\
		and\ \citenamefont {Date}}]{Takeuchi92b}%
	\BibitemOpen
	\bibfield  {author} {\bibinfo {author} {\bibfnamefont {T.}~\bibnamefont
			{Takeuchi}}, \bibinfo {author} {\bibfnamefont {M.}~\bibnamefont {Ono}},
		\bibinfo {author} {\bibfnamefont {H.}~\bibnamefont {Hori}}, \bibinfo {author}
		{\bibfnamefont {T.}~\bibnamefont {Yosida}}, \bibinfo {author} {\bibfnamefont
			{A.}~\bibnamefont {Yamagishi}}, \ and\ \bibinfo {author} {\bibfnamefont
			{M.}~\bibnamefont {Date}},\ }\href {\doibase 10.1143/JPSJ.61.3255} {\bibfield
		{journal} {\bibinfo  {journal} {Journal of the Physical Society of Japan}\
		}\textbf {\bibinfo {volume} {61}},\ \bibinfo {pages} {3255} (\bibinfo {year}
		{1992}{\natexlab{b}})}\BibitemShut {NoStop}%
	\bibitem [{\citenamefont {Meyer}\ \emph {et~al.}(1982)\citenamefont {Meyer},
		\citenamefont {Gleizes}, \citenamefont {Girerd}, \citenamefont {Verdaguer},\
		and\ \citenamefont {Kahn}}]{Meyer}%
	\BibitemOpen
	\bibfield  {author} {\bibinfo {author} {\bibfnamefont {A.}~\bibnamefont
			{Meyer}}, \bibinfo {author} {\bibfnamefont {A.}~\bibnamefont {Gleizes}},
		\bibinfo {author} {\bibfnamefont {J.~J.}\ \bibnamefont {Girerd}}, \bibinfo
		{author} {\bibfnamefont {M.}~\bibnamefont {Verdaguer}}, \ and\ \bibinfo
		{author} {\bibfnamefont {O.}~\bibnamefont {Kahn}},\ }\href {\doibase
		10.1021/ic00135a006} {\bibfield  {journal} {\bibinfo  {journal} {Inorganic
				Chemistry}\ }\textbf {\bibinfo {volume} {21}},\ \bibinfo {pages} {1729}
		(\bibinfo {year} {1982})}\BibitemShut {NoStop}%
	\bibitem [{\citenamefont {Honda}\ \emph {et~al.}(1998)\citenamefont {Honda},
		\citenamefont {Asakawa},\ and\ \citenamefont {Katsumata}}]{Honda}%
	\BibitemOpen
	\bibfield  {author} {\bibinfo {author} {\bibfnamefont {Z.}~\bibnamefont
			{Honda}}, \bibinfo {author} {\bibfnamefont {H.}~\bibnamefont {Asakawa}}, \
		and\ \bibinfo {author} {\bibfnamefont {K.}~\bibnamefont {Katsumata}},\ }\href
	{\doibase 10.1103/PhysRevLett.81.2566} {\bibfield  {journal} {\bibinfo
			{journal} {Physical Review Letters}\ }\textbf {\bibinfo {volume} {81}},\
		\bibinfo {pages} {2566} (\bibinfo {year} {1998})}\BibitemShut {NoStop}%
	\bibitem [{\citenamefont {Koike}\ \emph {et~al.}(2000)\citenamefont {Koike},
		\citenamefont {Metoki}, \citenamefont {Morii}, \citenamefont {Kobayashi},
		\citenamefont {Ishii},\ and\ \citenamefont {Yamashita}}]{Koike}%
	\BibitemOpen
	\bibfield  {author} {\bibinfo {author} {\bibfnamefont {Y.}~\bibnamefont
			{Koike}}, \bibinfo {author} {\bibfnamefont {N.}~\bibnamefont {Metoki}},
		\bibinfo {author} {\bibfnamefont {Y.}~\bibnamefont {Morii}}, \bibinfo
		{author} {\bibfnamefont {T.}~\bibnamefont {Kobayashi}}, \bibinfo {author}
		{\bibfnamefont {T.}~\bibnamefont {Ishii}}, \ and\ \bibinfo {author}
		{\bibfnamefont {M.}~\bibnamefont {Yamashita}},\ }\href {\doibase
		10.1143/JPSJ.69.4034} {\bibfield  {journal} {\bibinfo  {journal} {Journal of
				the Physical Society of Japan}\ }\textbf {\bibinfo {volume} {69}},\ \bibinfo
		{pages} {4034} (\bibinfo {year} {2000})}\BibitemShut {NoStop}%
	\bibitem [{\citenamefont {Johnson}\ \emph {et~al.}(1979)\citenamefont
		{Johnson}, \citenamefont {Rayne},\ and\ \citenamefont
		{Friedberg}}]{Johnson1979}%
	\BibitemOpen
	\bibfield  {author} {\bibinfo {author} {\bibfnamefont {P.~B.}\ \bibnamefont
			{Johnson}}, \bibinfo {author} {\bibfnamefont {J.~A.}\ \bibnamefont {Rayne}},
		\ and\ \bibinfo {author} {\bibfnamefont {S.~A.}\ \bibnamefont {Friedberg}},\
	}\href {\doibase 10.1063/1.327143} {\bibfield  {journal} {\bibinfo  {journal}
			{Journal of Applied Physics}\ }\textbf {\bibinfo {volume} {50}},\ \bibinfo
		{pages} {1853} (\bibinfo {year} {1979})}\BibitemShut {NoStop}%
	\bibitem [{\citenamefont {Buyers}\ \emph {et~al.}(1986)\citenamefont {Buyers},
		\citenamefont {Morra}, \citenamefont {Armstrong}, \citenamefont {Hogan},
		\citenamefont {Gerlach},\ and\ \citenamefont {Hirakawa}}]{Buyers1986}%
	\BibitemOpen
	\bibfield  {author} {\bibinfo {author} {\bibfnamefont {W.~J.~L.}\
			\bibnamefont {Buyers}}, \bibinfo {author} {\bibfnamefont {R.~M.}\
			\bibnamefont {Morra}}, \bibinfo {author} {\bibfnamefont {R.~L.}\ \bibnamefont
			{Armstrong}}, \bibinfo {author} {\bibfnamefont {M.~J.}\ \bibnamefont
			{Hogan}}, \bibinfo {author} {\bibfnamefont {P.}~\bibnamefont {Gerlach}}, \
		and\ \bibinfo {author} {\bibfnamefont {K.}~\bibnamefont {Hirakawa}},\ }\href
	{\doibase 10.1103/PhysRevLett.56.371} {\bibfield  {journal} {\bibinfo
			{journal} {Physical Review Letters}\ }\textbf {\bibinfo {volume} {56}},\
		\bibinfo {pages} {371} (\bibinfo {year} {1986})}\BibitemShut {NoStop}%
	\bibitem [{\citenamefont {Morra}\ \emph {et~al.}(1988)\citenamefont {Morra},
		\citenamefont {Buyers}, \citenamefont {Armstrong},\ and\ \citenamefont
		{Hirakawa}}]{Morra}%
	\BibitemOpen
	\bibfield  {author} {\bibinfo {author} {\bibfnamefont {R.~M.}\ \bibnamefont
			{Morra}}, \bibinfo {author} {\bibfnamefont {W.~J.~L.}\ \bibnamefont
			{Buyers}}, \bibinfo {author} {\bibfnamefont {R.~L.}\ \bibnamefont
			{Armstrong}}, \ and\ \bibinfo {author} {\bibfnamefont {K.}~\bibnamefont
			{Hirakawa}},\ }\href {\doibase 10.1103/PhysRevB.38.543} {\bibfield  {journal}
		{\bibinfo  {journal} {Physical Review B}\ }\textbf {\bibinfo {volume} {38}},\
		\bibinfo {pages} {543} (\bibinfo {year} {1988})}\BibitemShut {NoStop}%
	\bibitem [{\citenamefont {Brambleby}\ \emph {et~al.}(2017)\citenamefont
		{Brambleby}, \citenamefont {Manson}, \citenamefont {Goddard}, \citenamefont
		{Stone}, \citenamefont {Johnson}, \citenamefont {Manuel}, \citenamefont
		{Villa}, \citenamefont {Brown}, \citenamefont {Lu}, \citenamefont {Chikara},
		\citenamefont {Zapf}, \citenamefont {Lapidus}, \citenamefont {Scatena},
		\citenamefont {Macchi}, \citenamefont {Chen}, \citenamefont {Wu},\ and\
		\citenamefont {Singleton}}]{Brambley}%
	\BibitemOpen
	\bibfield  {author} {\bibinfo {author} {\bibfnamefont {J.}~\bibnamefont
			{Brambleby}}, \bibinfo {author} {\bibfnamefont {J.~L.}\ \bibnamefont
			{Manson}}, \bibinfo {author} {\bibfnamefont {P.~A.}\ \bibnamefont {Goddard}},
		\bibinfo {author} {\bibfnamefont {M.~B.}\ \bibnamefont {Stone}}, \bibinfo
		{author} {\bibfnamefont {R.~D.}\ \bibnamefont {Johnson}}, \bibinfo {author}
		{\bibfnamefont {P.}~\bibnamefont {Manuel}}, \bibinfo {author} {\bibfnamefont
			{J.~A.}\ \bibnamefont {Villa}}, \bibinfo {author} {\bibfnamefont {C.~M.}\
			\bibnamefont {Brown}}, \bibinfo {author} {\bibfnamefont {H.}~\bibnamefont
			{Lu}}, \bibinfo {author} {\bibfnamefont {S.}~\bibnamefont {Chikara}},
		\bibinfo {author} {\bibfnamefont {V.}~\bibnamefont {Zapf}}, \bibinfo {author}
		{\bibfnamefont {S.~H.}\ \bibnamefont {Lapidus}}, \bibinfo {author}
		{\bibfnamefont {R.}~\bibnamefont {Scatena}}, \bibinfo {author} {\bibfnamefont
			{P.}~\bibnamefont {Macchi}}, \bibinfo {author} {\bibfnamefont {Y.-s.}\
			\bibnamefont {Chen}}, \bibinfo {author} {\bibfnamefont {L.-C.}\ \bibnamefont
			{Wu}}, \ and\ \bibinfo {author} {\bibfnamefont {J.}~\bibnamefont
			{Singleton}},\ }\href {\doibase 10.1103/PhysRevB.95.134435} {\bibfield
		{journal} {\bibinfo  {journal} {Physical Review B}\ }\textbf {\bibinfo
			{volume} {95}},\ \bibinfo {pages} {134435} (\bibinfo {year}
		{2017})}\BibitemShut {NoStop}%
	\bibitem [{\citenamefont {Lipps}\ \emph {et~al.}(2017)\citenamefont {Lipps},
		\citenamefont {Arkenbout}, \citenamefont {Polyakov}, \citenamefont
		{Günther}, \citenamefont {Salikhov}, \citenamefont {Vavilova}, \citenamefont
		{Klauss}, \citenamefont {Büchner}, \citenamefont {Palstra},\ and\
		\citenamefont {Kataev}}]{Lipps2017}%
	\BibitemOpen
	\bibfield  {author} {\bibinfo {author} {\bibfnamefont {F.}~\bibnamefont
			{Lipps}}, \bibinfo {author} {\bibfnamefont {A.~H.}\ \bibnamefont
			{Arkenbout}}, \bibinfo {author} {\bibfnamefont {A.}~\bibnamefont {Polyakov}},
		\bibinfo {author} {\bibfnamefont {M.}~\bibnamefont {Günther}}, \bibinfo
		{author} {\bibfnamefont {T.}~\bibnamefont {Salikhov}}, \bibinfo {author}
		{\bibfnamefont {E.}~\bibnamefont {Vavilova}}, \bibinfo {author}
		{\bibfnamefont {H.~H.}\ \bibnamefont {Klauss}}, \bibinfo {author}
		{\bibfnamefont {B.}~\bibnamefont {Büchner}}, \bibinfo {author}
		{\bibfnamefont {T.~M.}\ \bibnamefont {Palstra}}, \ and\ \bibinfo {author}
		{\bibfnamefont {V.}~\bibnamefont {Kataev}},\ }\href {\doibase
		10.1063/1.5010316} {\bibfield  {journal} {\bibinfo  {journal} {Low
				Temperature Physics}\ }\textbf {\bibinfo {volume} {43}},\ \bibinfo {pages}
		{1298} (\bibinfo {year} {2017})}\BibitemShut {NoStop}%
	\bibitem [{\citenamefont {Vasiliev}\ \emph {et~al.}(2004)\citenamefont
		{Vasiliev}, \citenamefont {Ignatchik}, \citenamefont {Isobe},\ and\
		\citenamefont {Ueda}}]{Vasiliev}%
	\BibitemOpen
	\bibfield  {author} {\bibinfo {author} {\bibfnamefont {A.~N.}\ \bibnamefont
			{Vasiliev}}, \bibinfo {author} {\bibfnamefont {O.~L.}\ \bibnamefont
			{Ignatchik}}, \bibinfo {author} {\bibfnamefont {M.}~\bibnamefont {Isobe}}, \
		and\ \bibinfo {author} {\bibfnamefont {Y.}~\bibnamefont {Ueda}},\ }\href
	{\doibase 10.1103/PhysRevB.70.132415} {\bibfield  {journal} {\bibinfo
			{journal} {Physical Review B}\ }\textbf {\bibinfo {volume} {70}},\ \bibinfo
		{pages} {132415} (\bibinfo {year} {2004})}\BibitemShut {NoStop}%
	\bibitem [{\citenamefont {Law}\ \emph {et~al.}(2014)\citenamefont {Law},
		\citenamefont {Koo}, \citenamefont {Whangbo}, \citenamefont {Br\"ucher},
		\citenamefont {Pomjakushin},\ and\ \citenamefont {Kremer}}]{Law2014}%
	\BibitemOpen
	\bibfield  {author} {\bibinfo {author} {\bibfnamefont {J.~M.}\ \bibnamefont
			{Law}}, \bibinfo {author} {\bibfnamefont {H.-J.}\ \bibnamefont {Koo}},
		\bibinfo {author} {\bibfnamefont {M.-H.}\ \bibnamefont {Whangbo}}, \bibinfo
		{author} {\bibfnamefont {E.}~\bibnamefont {Br\"ucher}}, \bibinfo {author}
		{\bibfnamefont {V.}~\bibnamefont {Pomjakushin}}, \ and\ \bibinfo {author}
		{\bibfnamefont {R.~K.}\ \bibnamefont {Kremer}},\ }\href {\doibase
		10.1103/PhysRevB.89.014423} {\bibfield  {journal} {\bibinfo  {journal}
			{Physical Review B}\ }\textbf {\bibinfo {volume} {89}},\ \bibinfo {pages}
		{014423} (\bibinfo {year} {2014})}\BibitemShut {NoStop}%
	\bibitem [{\citenamefont {Christian}\ \emph {et~al.}(2017)\citenamefont
		{Christian}, \citenamefont {Hunt},\ and\ \citenamefont
		{Neumeier}}]{Christian}%
	\BibitemOpen
	\bibfield  {author} {\bibinfo {author} {\bibfnamefont {A.~B.}\ \bibnamefont
			{Christian}}, \bibinfo {author} {\bibfnamefont {C.~D.}\ \bibnamefont {Hunt}},
		\ and\ \bibinfo {author} {\bibfnamefont {J.~J.}\ \bibnamefont {Neumeier}},\
	}\href {\doibase 10.1103/PhysRevB.96.024433} {\bibfield  {journal} {\bibinfo
			{journal} {Physical Review B}\ }\textbf {\bibinfo {volume} {96}},\ \bibinfo
		{pages} {024433} (\bibinfo {year} {2017})}\BibitemShut {NoStop}%
	\bibitem [{\citenamefont {Williams}\ \emph {et~al.}(2020)\citenamefont
		{Williams}, \citenamefont {Blackmore}, \citenamefont {Curley}, \citenamefont
		{Lees}, \citenamefont {Birnbaum}, \citenamefont {Singleton}, \citenamefont
		{Huddart}, \citenamefont {Hicken}, \citenamefont {Lancaster}, \citenamefont
		{Blundell}, \citenamefont {Xiao}, \citenamefont {Ozarowski}, \citenamefont
		{Pratt}, \citenamefont {Voneshen}, \citenamefont {Guguchia}, \citenamefont
		{Baines}, \citenamefont {Schlueter}, \citenamefont {Villa}, \citenamefont
		{Manson},\ and\ \citenamefont {Goddard}}]{Williams2020}%
	\BibitemOpen
	\bibfield  {author} {\bibinfo {author} {\bibfnamefont {R.~C.}\ \bibnamefont
			{Williams}}, \bibinfo {author} {\bibfnamefont {W.~J.~A.}\ \bibnamefont
			{Blackmore}}, \bibinfo {author} {\bibfnamefont {S.~P.~M.}\ \bibnamefont
			{Curley}}, \bibinfo {author} {\bibfnamefont {M.~R.}\ \bibnamefont {Lees}},
		\bibinfo {author} {\bibfnamefont {S.~M.}\ \bibnamefont {Birnbaum}}, \bibinfo
		{author} {\bibfnamefont {J.}~\bibnamefont {Singleton}}, \bibinfo {author}
		{\bibfnamefont {B.~M.}\ \bibnamefont {Huddart}}, \bibinfo {author}
		{\bibfnamefont {T.~J.}\ \bibnamefont {Hicken}}, \bibinfo {author}
		{\bibfnamefont {T.}~\bibnamefont {Lancaster}}, \bibinfo {author}
		{\bibfnamefont {S.~J.}\ \bibnamefont {Blundell}}, \bibinfo {author}
		{\bibfnamefont {F.}~\bibnamefont {Xiao}}, \bibinfo {author} {\bibfnamefont
			{A.}~\bibnamefont {Ozarowski}}, \bibinfo {author} {\bibfnamefont {F.~L.}\
			\bibnamefont {Pratt}}, \bibinfo {author} {\bibfnamefont {D.~J.}\ \bibnamefont
			{Voneshen}}, \bibinfo {author} {\bibfnamefont {Z.}~\bibnamefont {Guguchia}},
		\bibinfo {author} {\bibfnamefont {C.}~\bibnamefont {Baines}}, \bibinfo
		{author} {\bibfnamefont {J.~A.}\ \bibnamefont {Schlueter}}, \bibinfo {author}
		{\bibfnamefont {D.~Y.}\ \bibnamefont {Villa}}, \bibinfo {author}
		{\bibfnamefont {J.~L.}\ \bibnamefont {Manson}}, \ and\ \bibinfo {author}
		{\bibfnamefont {P.~A.}\ \bibnamefont {Goddard}},\ }\href {\doibase
		10.1103/PhysRevResearch.2.013082} {\bibfield  {journal} {\bibinfo  {journal}
			{Physical Review Research}\ }\textbf {\bibinfo {volume} {2}},\ \bibinfo
		{pages} {013082} (\bibinfo {year} {2020})}\BibitemShut {NoStop}%
	\bibitem [{\citenamefont {Zheludev}\ \emph {et~al.}(2000)\citenamefont
		{Zheludev}, \citenamefont {Masuda}, \citenamefont {Tsukada}, \citenamefont
		{Uchiyama}, \citenamefont {Uchinokura}, \citenamefont {B\"oni},\ and\
		\citenamefont {Lee}}]{Zheludev2000}%
	\BibitemOpen
	\bibfield  {author} {\bibinfo {author} {\bibfnamefont {A.}~\bibnamefont
			{Zheludev}}, \bibinfo {author} {\bibfnamefont {T.}~\bibnamefont {Masuda}},
		\bibinfo {author} {\bibfnamefont {I.}~\bibnamefont {Tsukada}}, \bibinfo
		{author} {\bibfnamefont {Y.}~\bibnamefont {Uchiyama}}, \bibinfo {author}
		{\bibfnamefont {K.}~\bibnamefont {Uchinokura}}, \bibinfo {author}
		{\bibfnamefont {P.}~\bibnamefont {B\"oni}}, \ and\ \bibinfo {author}
		{\bibfnamefont {S.-H.}\ \bibnamefont {Lee}},\ }\href {\doibase
		10.1103/PhysRevB.62.8921} {\bibfield  {journal} {\bibinfo  {journal}
			{Physical Review B}\ }\textbf {\bibinfo {volume} {62}},\ \bibinfo {pages}
		{8921} (\bibinfo {year} {2000})}\BibitemShut {NoStop}%
	\bibitem [{\citenamefont {Smirnov}\ \emph {et~al.}(2002)\citenamefont
		{Smirnov}, \citenamefont {Glazkov}, \citenamefont {Krug~von Nidda},
		\citenamefont {Loidl}, \citenamefont {Demianets},\ and\ \citenamefont
		{Shapiro}}]{Smirnov2002}%
	\BibitemOpen
	\bibfield  {author} {\bibinfo {author} {\bibfnamefont {A.~I.}\ \bibnamefont
			{Smirnov}}, \bibinfo {author} {\bibfnamefont {V.~N.}\ \bibnamefont
			{Glazkov}}, \bibinfo {author} {\bibfnamefont {H.-A.}\ \bibnamefont {Krug~von
				Nidda}}, \bibinfo {author} {\bibfnamefont {A.}~\bibnamefont {Loidl}},
		\bibinfo {author} {\bibfnamefont {L.~N.}\ \bibnamefont {Demianets}}, \ and\
		\bibinfo {author} {\bibfnamefont {A.~Y.}\ \bibnamefont {Shapiro}},\ }\href
	{\doibase 10.1103/PhysRevB.65.174422} {\bibfield  {journal} {\bibinfo
			{journal} {Physical Review B}\ }\textbf {\bibinfo {volume} {65}},\ \bibinfo
		{pages} {174422} (\bibinfo {year} {2002})}\BibitemShut {NoStop}%
	\bibitem [{\citenamefont {Tsujii}\ \emph {et~al.}(2005)\citenamefont {Tsujii},
		\citenamefont {Suzuki}, \citenamefont {Suzuki}, \citenamefont {Kitazawa},\
		and\ \citenamefont {Kido}}]{Tsujii2005b}%
	\BibitemOpen
	\bibfield  {author} {\bibinfo {author} {\bibfnamefont {N.}~\bibnamefont
			{Tsujii}}, \bibinfo {author} {\bibfnamefont {O.}~\bibnamefont {Suzuki}},
		\bibinfo {author} {\bibfnamefont {H.}~\bibnamefont {Suzuki}}, \bibinfo
		{author} {\bibfnamefont {H.}~\bibnamefont {Kitazawa}}, \ and\ \bibinfo
		{author} {\bibfnamefont {G.}~\bibnamefont {Kido}},\ }\href {\doibase
		10.1103/PhysRevB.72.104402} {\bibfield  {journal} {\bibinfo  {journal}
			{Physical Review B}\ }\textbf {\bibinfo {volume} {72}},\ \bibinfo {pages}
		{104402} (\bibinfo {year} {2005})}\BibitemShut {NoStop}%
	\bibitem [{\citenamefont {Pahari}\ \emph {et~al.}(2008)\citenamefont {Pahari},
		\citenamefont {Ghoshray}, \citenamefont {Sarkar},\ and\ \citenamefont
		{Ghoshray}}]{Pahari}%
	\BibitemOpen
	\bibfield  {author} {\bibinfo {author} {\bibfnamefont {B.}~\bibnamefont
			{Pahari}}, \bibinfo {author} {\bibfnamefont {K.}~\bibnamefont {Ghoshray}},
		\bibinfo {author} {\bibfnamefont {R.}~\bibnamefont {Sarkar}}, \ and\ \bibinfo
		{author} {\bibfnamefont {A.}~\bibnamefont {Ghoshray}},\ }\href {\doibase
		10.1103/PhysRevB.77.224429} {\bibfield  {journal} {\bibinfo  {journal}
			{Physical Review B}\ }\textbf {\bibinfo {volume} {77}},\ \bibinfo {pages}
		{224429} (\bibinfo {year} {2008})}\BibitemShut {NoStop}%
	\bibitem [{\citenamefont {Bera}\ \emph {et~al.}(2013)\citenamefont {Bera},
		\citenamefont {Lake}, \citenamefont {Islam}, \citenamefont {Klemke},
		\citenamefont {Faulhaber},\ and\ \citenamefont {Law}}]{Bera}%
	\BibitemOpen
	\bibfield  {author} {\bibinfo {author} {\bibfnamefont {A.~K.}\ \bibnamefont
			{Bera}}, \bibinfo {author} {\bibfnamefont {B.}~\bibnamefont {Lake}}, \bibinfo
		{author} {\bibfnamefont {A.~T. M.~N.}\ \bibnamefont {Islam}}, \bibinfo
		{author} {\bibfnamefont {B.}~\bibnamefont {Klemke}}, \bibinfo {author}
		{\bibfnamefont {E.}~\bibnamefont {Faulhaber}}, \ and\ \bibinfo {author}
		{\bibfnamefont {J.~M.}\ \bibnamefont {Law}},\ }\href {\doibase
		10.1103/PhysRevB.87.224423} {\bibfield  {journal} {\bibinfo  {journal}
			{Physical Review B}\ }\textbf {\bibinfo {volume} {87}},\ \bibinfo {pages}
		{224423} (\bibinfo {year} {2013})}\BibitemShut {NoStop}%
	\bibitem [{\citenamefont {Tin}\ \emph {et~al.}(2023)\citenamefont {Tin},
		\citenamefont {Jenkins}, \citenamefont {Xing}, \citenamefont {Caci},
		\citenamefont {Gai}, \citenamefont {Jin}, \citenamefont {Wessel},
		\citenamefont {Krzystek}, \citenamefont {Li}, \citenamefont {Daemen},
		\citenamefont {Cheng},\ and\ \citenamefont {Xue}}]{TinNatComm}%
	\BibitemOpen
	\bibfield  {author} {\bibinfo {author} {\bibfnamefont {P.}~\bibnamefont
			{Tin}}, \bibinfo {author} {\bibfnamefont {M.~J.}\ \bibnamefont {Jenkins}},
		\bibinfo {author} {\bibfnamefont {J.}~\bibnamefont {Xing}}, \bibinfo {author}
		{\bibfnamefont {N.}~\bibnamefont {Caci}}, \bibinfo {author} {\bibfnamefont
			{Z.}~\bibnamefont {Gai}}, \bibinfo {author} {\bibfnamefont {R.}~\bibnamefont
			{Jin}}, \bibinfo {author} {\bibfnamefont {S.}~\bibnamefont {Wessel}},
		\bibinfo {author} {\bibfnamefont {J.}~\bibnamefont {Krzystek}}, \bibinfo
		{author} {\bibfnamefont {C.}~\bibnamefont {Li}}, \bibinfo {author}
		{\bibfnamefont {L.~L.}\ \bibnamefont {Daemen}}, \bibinfo {author}
		{\bibfnamefont {Y.}~\bibnamefont {Cheng}}, \ and\ \bibinfo {author}
		{\bibfnamefont {Z.-L.}\ \bibnamefont {Xue}},\ }\href {\doibase
		10.1038/s41467-023-41014-1} {\bibfield  {journal} {\bibinfo  {journal}
			{Nature Communications}\ }\textbf {\bibinfo {volume} {14}},\ \bibinfo {pages}
		{5454} (\bibinfo {year} {2023})}\BibitemShut {NoStop}%
	\bibitem [{\citenamefont {Čižmár}\ \emph {et~al.}(2008)\citenamefont
		{Čižmár}, \citenamefont {Ozerov}, \citenamefont {Ignatchik}, \citenamefont
		{Papageorgiou}, \citenamefont {Wosnitza}, \citenamefont {Zvyagin},
		\citenamefont {Krzystek}, \citenamefont {Zhou}, \citenamefont {Landee},
		\citenamefont {Landry}, \citenamefont {Turnbull},\ and\ \citenamefont
		{Wikaira}}]{Cizmar2008}%
	\BibitemOpen
	\bibfield  {author} {\bibinfo {author} {\bibfnamefont {E.}~\bibnamefont
			{Čižmár}}, \bibinfo {author} {\bibfnamefont {M.}~\bibnamefont {Ozerov}},
		\bibinfo {author} {\bibfnamefont {O.}~\bibnamefont {Ignatchik}}, \bibinfo
		{author} {\bibfnamefont {T.~P.}\ \bibnamefont {Papageorgiou}}, \bibinfo
		{author} {\bibfnamefont {J.}~\bibnamefont {Wosnitza}}, \bibinfo {author}
		{\bibfnamefont {S.~A.}\ \bibnamefont {Zvyagin}}, \bibinfo {author}
		{\bibfnamefont {J.}~\bibnamefont {Krzystek}}, \bibinfo {author}
		{\bibfnamefont {Z.}~\bibnamefont {Zhou}}, \bibinfo {author} {\bibfnamefont
			{C.~P.}\ \bibnamefont {Landee}}, \bibinfo {author} {\bibfnamefont {B.~R.}\
			\bibnamefont {Landry}}, \bibinfo {author} {\bibfnamefont {M.~M.}\
			\bibnamefont {Turnbull}}, \ and\ \bibinfo {author} {\bibfnamefont {J.~L.}\
			\bibnamefont {Wikaira}},\ }\href {\doibase 10.1088/1367-2630/10/3/033008}
	{\bibfield  {journal} {\bibinfo  {journal} {New Journal of Physics}\ }\textbf
		{\bibinfo {volume} {10}},\ \bibinfo {pages} {033008} (\bibinfo {year}
		{2008})}\BibitemShut {NoStop}%
	\bibitem [{\citenamefont {Li}\ \emph {et~al.}(2025)\citenamefont {Li},
		\citenamefont {Wu}, \citenamefont {Wang}, \citenamefont {Luo}, \citenamefont
		{Du}, \citenamefont {Xu}, \citenamefont {Hu}, \citenamefont {Chen},
		\citenamefont {Yang}, \citenamefont {Liu}, \citenamefont {Yu}, \citenamefont
		{Cui}, \citenamefont {Zhou}, \citenamefont {Lu},\ and\ \citenamefont
		{Yu}}]{Li2025}%
	\BibitemOpen
	\bibfield  {author} {\bibinfo {author} {\bibfnamefont {S.}~\bibnamefont
			{Li}}, \bibinfo {author} {\bibfnamefont {Z.}~\bibnamefont {Wu}}, \bibinfo
		{author} {\bibfnamefont {Y.}~\bibnamefont {Wang}}, \bibinfo {author}
		{\bibfnamefont {J.}~\bibnamefont {Luo}}, \bibinfo {author} {\bibfnamefont
			{K.}~\bibnamefont {Du}}, \bibinfo {author} {\bibfnamefont {X.}~\bibnamefont
			{Xu}}, \bibinfo {author} {\bibfnamefont {Z.}~\bibnamefont {Hu}}, \bibinfo
		{author} {\bibfnamefont {Y.}~\bibnamefont {Chen}}, \bibinfo {author}
		{\bibfnamefont {J.}~\bibnamefont {Yang}}, \bibinfo {author} {\bibfnamefont
			{Z.}~\bibnamefont {Liu}}, \bibinfo {author} {\bibfnamefont {R.}~\bibnamefont
			{Yu}}, \bibinfo {author} {\bibfnamefont {Y.}~\bibnamefont {Cui}}, \bibinfo
		{author} {\bibfnamefont {R.}~\bibnamefont {Zhou}}, \bibinfo {author}
		{\bibfnamefont {H.}~\bibnamefont {Lu}}, \ and\ \bibinfo {author}
		{\bibfnamefont {W.}~\bibnamefont {Yu}},\ }\href {\doibase
		10.1103/PhysRevB.111.195164} {\bibfield  {journal} {\bibinfo  {journal}
			{Physical Review B}\ }\textbf {\bibinfo {volume} {111}},\ \bibinfo {pages}
		{195164} (\bibinfo {year} {2025})}\BibitemShut {NoStop}%
	\bibitem [{\citenamefont {Chou}\ \emph {et~al.}(1994)\citenamefont {Chou},
		\citenamefont {Abboud}, \citenamefont {Talham}, \citenamefont {Kim},\ and\
		\citenamefont {Meisel}}]{Chou}%
	\BibitemOpen
	\bibfield  {author} {\bibinfo {author} {\bibfnamefont {L.-K.}\ \bibnamefont
			{Chou}}, \bibinfo {author} {\bibfnamefont {K.~A.}\ \bibnamefont {Abboud}},
		\bibinfo {author} {\bibfnamefont {D.~R.}\ \bibnamefont {Talham}}, \bibinfo
		{author} {\bibfnamefont {W.~W.}\ \bibnamefont {Kim}}, \ and\ \bibinfo
		{author} {\bibfnamefont {M.~W.}\ \bibnamefont {Meisel}},\ }\href {\doibase
		10.1021/cm00047a026} {\bibfield  {journal} {\bibinfo  {journal} {Chemistry of
				Materials}\ }\textbf {\bibinfo {volume} {6}},\ \bibinfo {pages} {2051}
		(\bibinfo {year} {1994})}\BibitemShut {NoStop}%
	\bibitem [{\citenamefont {Goto}\ \emph {et~al.}(2006)\citenamefont {Goto},
		\citenamefont {Ishikawa}, \citenamefont {Shimaoka},\ and\ \citenamefont
		{Fujii}}]{GotoPRB}%
	\BibitemOpen
	\bibfield  {author} {\bibinfo {author} {\bibfnamefont {T.}~\bibnamefont
			{Goto}}, \bibinfo {author} {\bibfnamefont {T.}~\bibnamefont {Ishikawa}},
		\bibinfo {author} {\bibfnamefont {Y.}~\bibnamefont {Shimaoka}}, \ and\
		\bibinfo {author} {\bibfnamefont {Y.}~\bibnamefont {Fujii}},\ }\href
	{\doibase 10.1103/PhysRevB.73.214406} {\bibfield  {journal} {\bibinfo
			{journal} {Physical Review B}\ }\textbf {\bibinfo {volume} {73}},\ \bibinfo
		{pages} {214406} (\bibinfo {year} {2006})}\BibitemShut {NoStop}%
	\bibitem [{\citenamefont {Koido}\ \emph {et~al.}(1995)\citenamefont {Koido},
		\citenamefont {Kambe}, \citenamefont {Teraoka}, \citenamefont {Hirai},
		\citenamefont {Goto}, \citenamefont {Chou}, \citenamefont {Talham},
		\citenamefont {Meisel},\ and\ \citenamefont {Nagata}}]{Koido}%
	\BibitemOpen
	\bibfield  {author} {\bibinfo {author} {\bibfnamefont {N.}~\bibnamefont
			{Koido}}, \bibinfo {author} {\bibfnamefont {T.}~\bibnamefont {Kambe}},
		\bibinfo {author} {\bibfnamefont {S.}~\bibnamefont {Teraoka}}, \bibinfo
		{author} {\bibfnamefont {S.}~\bibnamefont {Hirai}}, \bibinfo {author}
		{\bibfnamefont {T.}~\bibnamefont {Goto}}, \bibinfo {author} {\bibfnamefont
			{L.-K.}\ \bibnamefont {Chou}}, \bibinfo {author} {\bibfnamefont
			{D.}~\bibnamefont {Talham}}, \bibinfo {author} {\bibfnamefont
			{M.}~\bibnamefont {Meisel}}, \ and\ \bibinfo {author} {\bibfnamefont
			{K.}~\bibnamefont {Nagata}},\ }\href {\doibase
		https://doi.org/10.1016/0304-8853(94)01097-8} {\bibfield  {journal} {\bibinfo
			{journal} {Journal of Magnetism and Magnetic Materials}\ }\textbf {\bibinfo
			{volume} {140-144}},\ \bibinfo {pages} {1639} (\bibinfo {year} {1995})},\
	\bibinfo {note} {international Conference on Magnetism}\BibitemShut {NoStop}%
	\bibitem [{\citenamefont {{\v{C}}i{\v{z}}m{\'a}r}\ \emph
		{et~al.}(2002)\citenamefont {{\v{C}}i{\v{z}}m{\'a}r}, \citenamefont
		{Orend{\'a}{\v{c}}}, \citenamefont {Feher},\ and\ \citenamefont
		{Meisel}}]{Cizmar2002}%
	\BibitemOpen
	\bibfield  {author} {\bibinfo {author} {\bibfnamefont {E.}~\bibnamefont
			{{\v{C}}i{\v{z}}m{\'a}r}}, \bibinfo {author} {\bibfnamefont {M.}~\bibnamefont
			{Orend{\'a}{\v{c}}}}, \bibinfo {author} {\bibfnamefont {A.}~\bibnamefont
			{Feher}}, \ and\ \bibinfo {author} {\bibfnamefont {M.~W.}\ \bibnamefont
			{Meisel}},\ }\href {\doibase 10.1007/s10582-002-0071-7} {\bibfield  {journal}
		{\bibinfo  {journal} {Czechoslovak Journal of Physics}\ }\textbf {\bibinfo
			{volume} {52}},\ \bibinfo {pages} {A293} (\bibinfo {year}
		{2002})}\BibitemShut {NoStop}%
	\bibitem [{\citenamefont {Konik}\ and\ \citenamefont
		{Fendley}(2002)}]{Konik2002}%
	\BibitemOpen
	\bibfield  {author} {\bibinfo {author} {\bibfnamefont {R.~M.}\ \bibnamefont
			{Konik}}\ and\ \bibinfo {author} {\bibfnamefont {P.}~\bibnamefont
			{Fendley}},\ }\href {\doibase 10.1103/PhysRevB.66.144416} {\bibfield
		{journal} {\bibinfo  {journal} {Physical Review B}\ }\textbf {\bibinfo
			{volume} {66}},\ \bibinfo {pages} {144416} (\bibinfo {year}
		{2002})}\BibitemShut {NoStop}%
	\bibitem [{\citenamefont {F\'ath}(2003)}]{Fath03}%
	\BibitemOpen
	\bibfield  {author} {\bibinfo {author} {\bibfnamefont {G.}~\bibnamefont
			{F\'ath}},\ }\href {\doibase 10.1103/PhysRevB.68.134445} {\bibfield
		{journal} {\bibinfo  {journal} {Physical Review B}\ }\textbf {\bibinfo
			{volume} {68}},\ \bibinfo {pages} {134445} (\bibinfo {year}
		{2003})}\BibitemShut {NoStop}%
	\bibitem [{\citenamefont {Affleck}(1990)}]{Affleck1990}%
	\BibitemOpen
	\bibfield  {author} {\bibinfo {author} {\bibfnamefont {I.}~\bibnamefont
			{Affleck}},\ }\href {\doibase 10.1103/PhysRevB.41.6697} {\bibfield  {journal}
		{\bibinfo  {journal} {Physical Review B}\ }\textbf {\bibinfo {volume} {41}},\
		\bibinfo {pages} {6697} (\bibinfo {year} {1990})}\BibitemShut {NoStop}%
	\bibitem [{\citenamefont {Affleck}(1991)}]{Affleck1991}%
	\BibitemOpen
	\bibfield  {author} {\bibinfo {author} {\bibfnamefont {I.}~\bibnamefont
			{Affleck}},\ }\href {\doibase 10.1103/PhysRevB.43.3215} {\bibfield  {journal}
		{\bibinfo  {journal} {Physical Review B}\ }\textbf {\bibinfo {volume} {43}},\
		\bibinfo {pages} {3215} (\bibinfo {year} {1991})}\BibitemShut {NoStop}%
	\bibitem [{\citenamefont {Jakovac}\ \emph {et~al.}(2025)\citenamefont
		{Jakovac}, \citenamefont {Cvitani\'{c}}, \citenamefont {Ar\v{c}on},
		\citenamefont {Herak}, \citenamefont {Cin\v{c}i\'{c}}, \citenamefont
		{Topi\'{c}}, \citenamefont {Hosokoshi}, \citenamefont {Ono}, \citenamefont
		{Iwashita}, \citenamefont {Hayashi}, \citenamefont {Amaya}, \citenamefont
		{Matsuo}, \citenamefont {Kindo}, \citenamefont {Lon\v{c}ari\'{c}},
		\citenamefont {Horvati\'{c}}, \citenamefont {Takigawa},\ and\ \citenamefont
		{Grbi\'{c}}}]{Jakovac2024}%
	\BibitemOpen
	\bibfield  {author} {\bibinfo {author} {\bibfnamefont {I.}~\bibnamefont
			{Jakovac}}, \bibinfo {author} {\bibfnamefont {T.}~\bibnamefont
			{Cvitani\'{c}}}, \bibinfo {author} {\bibfnamefont {D.}~\bibnamefont
			{Ar\v{c}on}}, \bibinfo {author} {\bibfnamefont {M.}~\bibnamefont {Herak}},
		\bibinfo {author} {\bibfnamefont {D.}~\bibnamefont {Cin\v{c}i\'{c}}},
		\bibinfo {author} {\bibfnamefont {N.~B.}\ \bibnamefont {Topi\'{c}}}, \bibinfo
		{author} {\bibfnamefont {Y.}~\bibnamefont {Hosokoshi}}, \bibinfo {author}
		{\bibfnamefont {T.}~\bibnamefont {Ono}}, \bibinfo {author} {\bibfnamefont
			{K.}~\bibnamefont {Iwashita}}, \bibinfo {author} {\bibfnamefont
			{N.}~\bibnamefont {Hayashi}}, \bibinfo {author} {\bibfnamefont
			{N.}~\bibnamefont {Amaya}}, \bibinfo {author} {\bibfnamefont
			{A.}~\bibnamefont {Matsuo}}, \bibinfo {author} {\bibfnamefont
			{K.}~\bibnamefont {Kindo}}, \bibinfo {author} {\bibfnamefont
			{I.}~\bibnamefont {Lon\v{c}ari\'{c}}}, \bibinfo {author} {\bibfnamefont
			{M.}~\bibnamefont {Horvati\'{c}}}, \bibinfo {author} {\bibfnamefont
			{M.}~\bibnamefont {Takigawa}}, \ and\ \bibinfo {author} {\bibfnamefont
			{M.~S.}\ \bibnamefont {Grbi\'{c}}},\ }\href {\doibase
		10.1103/PhysRevB.111.064407} {\bibfield  {journal} {\bibinfo  {journal}
			{Physical Review B}\ }\textbf {\bibinfo {volume} {111}},\ \bibinfo {pages}
		{064407} (\bibinfo {year} {2025})}\BibitemShut {NoStop}%
	\bibitem [{\citenamefont {Nikuni}\ \emph {et~al.}(2000)\citenamefont {Nikuni},
		\citenamefont {Oshikawa}, \citenamefont {Oosawa},\ and\ \citenamefont
		{Tanaka}}]{Nikuni}%
	\BibitemOpen
	\bibfield  {author} {\bibinfo {author} {\bibfnamefont {T.}~\bibnamefont
			{Nikuni}}, \bibinfo {author} {\bibfnamefont {M.}~\bibnamefont {Oshikawa}},
		\bibinfo {author} {\bibfnamefont {A.}~\bibnamefont {Oosawa}}, \ and\ \bibinfo
		{author} {\bibfnamefont {H.}~\bibnamefont {Tanaka}},\ }\href {\doibase
		10.1103/PhysRevLett.84.5868} {\bibfield  {journal} {\bibinfo  {journal}
			{Physical Review Letters}\ }\textbf {\bibinfo {volume} {84}},\ \bibinfo
		{pages} {5868} (\bibinfo {year} {2000})}\BibitemShut {NoStop}%
	\bibitem [{\citenamefont {Campostrini}\ \emph {et~al.}(2001)\citenamefont
		{Campostrini}, \citenamefont {Hasenbusch}, \citenamefont {Pelissetto},
		\citenamefont {Rossi},\ and\ \citenamefont {Vicari}}]{Campostrini}%
	\BibitemOpen
	\bibfield  {author} {\bibinfo {author} {\bibfnamefont {M.}~\bibnamefont
			{Campostrini}}, \bibinfo {author} {\bibfnamefont {M.}~\bibnamefont
			{Hasenbusch}}, \bibinfo {author} {\bibfnamefont {A.}~\bibnamefont
			{Pelissetto}}, \bibinfo {author} {\bibfnamefont {P.}~\bibnamefont {Rossi}}, \
		and\ \bibinfo {author} {\bibfnamefont {E.}~\bibnamefont {Vicari}},\ }\href
	{\doibase 10.1103/PhysRevB.63.214503} {\bibfield  {journal} {\bibinfo
			{journal} {Physical Review B}\ }\textbf {\bibinfo {volume} {63}},\ \bibinfo
		{pages} {214503} (\bibinfo {year} {2001})}\BibitemShut {NoStop}%
	\bibitem [{\citenamefont {Jeong}\ \emph {et~al.}(2017)\citenamefont {Jeong},
		\citenamefont {Mayaffre}, \citenamefont {Berthier}, \citenamefont
		{Schmidiger}, \citenamefont {Zheludev},\ and\ \citenamefont
		{Horvatić}}]{Jeong2017}%
	\BibitemOpen
	\bibfield  {author} {\bibinfo {author} {\bibfnamefont {M.}~\bibnamefont
			{Jeong}}, \bibinfo {author} {\bibfnamefont {H.}~\bibnamefont {Mayaffre}},
		\bibinfo {author} {\bibfnamefont {C.}~\bibnamefont {Berthier}}, \bibinfo
		{author} {\bibfnamefont {D.}~\bibnamefont {Schmidiger}}, \bibinfo {author}
		{\bibfnamefont {A.}~\bibnamefont {Zheludev}}, \ and\ \bibinfo {author}
		{\bibfnamefont {M.}~\bibnamefont {Horvatić}},\ }\href {\doibase
		10.1103/PhysRevLett.118.167206} {\bibfield  {journal} {\bibinfo  {journal}
			{Physical Review Letters}\ }\textbf {\bibinfo {volume} {118}},\ \bibinfo
		{pages} {167206} (\bibinfo {year} {2017})}\BibitemShut {NoStop}%
	\bibitem [{\citenamefont {Blosser}\ \emph {et~al.}(2019)\citenamefont
		{Blosser}, \citenamefont {Horvati\ifmmode~\acute{c}\else \'{c}\fi{}},
		\citenamefont {Bewley}, \citenamefont {Gvasaliya},\ and\ \citenamefont
		{Zheludev}}]{Blosser}%
	\BibitemOpen
	\bibfield  {author} {\bibinfo {author} {\bibfnamefont {D.}~\bibnamefont
			{Blosser}}, \bibinfo {author} {\bibfnamefont {M.}~\bibnamefont
			{Horvati\ifmmode~\acute{c}\else \'{c}\fi{}}}, \bibinfo {author}
		{\bibfnamefont {R.}~\bibnamefont {Bewley}}, \bibinfo {author} {\bibfnamefont
			{S.}~\bibnamefont {Gvasaliya}}, \ and\ \bibinfo {author} {\bibfnamefont
			{A.}~\bibnamefont {Zheludev}},\ }\href {\doibase
		10.1103/PhysRevMaterials.3.074410} {\bibfield  {journal} {\bibinfo  {journal}
			{Physical Review Materials}\ }\textbf {\bibinfo {volume} {3}},\ \bibinfo
		{pages} {074410} (\bibinfo {year} {2019})}\BibitemShut {NoStop}%
	\bibitem [{\citenamefont {Jeong}\ \emph {et~al.}(2016)\citenamefont {Jeong},
		\citenamefont {Schmidiger}, \citenamefont {Mayaffre}, \citenamefont
		{Klanjšek}, \citenamefont {Berthier}, \citenamefont {Knafo}, \citenamefont
		{Ballon}, \citenamefont {Vignolle}, \citenamefont {Krämer}, \citenamefont
		{Zheludev},\ and\ \citenamefont {Horvatić}}]{Jeong2016}%
	\BibitemOpen
	\bibfield  {author} {\bibinfo {author} {\bibfnamefont {M.}~\bibnamefont
			{Jeong}}, \bibinfo {author} {\bibfnamefont {D.}~\bibnamefont {Schmidiger}},
		\bibinfo {author} {\bibfnamefont {H.}~\bibnamefont {Mayaffre}}, \bibinfo
		{author} {\bibfnamefont {M.}~\bibnamefont {Klanjšek}}, \bibinfo {author}
		{\bibfnamefont {C.}~\bibnamefont {Berthier}}, \bibinfo {author}
		{\bibfnamefont {W.}~\bibnamefont {Knafo}}, \bibinfo {author} {\bibfnamefont
			{G.}~\bibnamefont {Ballon}}, \bibinfo {author} {\bibfnamefont
			{B.}~\bibnamefont {Vignolle}}, \bibinfo {author} {\bibfnamefont
			{S.}~\bibnamefont {Krämer}}, \bibinfo {author} {\bibfnamefont
			{A.}~\bibnamefont {Zheludev}}, \ and\ \bibinfo {author} {\bibfnamefont
			{M.}~\bibnamefont {Horvatić}},\ }\href {\doibase
		10.1103/PhysRevLett.117.106402} {\bibfield  {journal} {\bibinfo  {journal}
			{Physical Review Letters}\ }\textbf {\bibinfo {volume} {117}},\ \bibinfo
		{pages} {106402} (\bibinfo {year} {2016})}\BibitemShut {NoStop}%
	\bibitem [{\citenamefont {Mukhopadhyay}\ \emph {et~al.}(2012)\citenamefont
		{Mukhopadhyay}, \citenamefont {Klanjšek}, \citenamefont {Grbić},
		\citenamefont {Blinder}, \citenamefont {Mayaffre}, \citenamefont {Berthier},
		\citenamefont {Horvatić}, \citenamefont {Continentino}, \citenamefont
		{Paduan-Filho}, \citenamefont {Chiari},\ and\ \citenamefont
		{Piovesana}}]{Mukhopadhyay2012}%
	\BibitemOpen
	\bibfield  {author} {\bibinfo {author} {\bibfnamefont {S.}~\bibnamefont
			{Mukhopadhyay}}, \bibinfo {author} {\bibfnamefont {M.}~\bibnamefont
			{Klanjšek}}, \bibinfo {author} {\bibfnamefont {M.~S.}\ \bibnamefont
			{Grbić}}, \bibinfo {author} {\bibfnamefont {R.}~\bibnamefont {Blinder}},
		\bibinfo {author} {\bibfnamefont {H.}~\bibnamefont {Mayaffre}}, \bibinfo
		{author} {\bibfnamefont {C.}~\bibnamefont {Berthier}}, \bibinfo {author}
		{\bibfnamefont {M.}~\bibnamefont {Horvatić}}, \bibinfo {author}
		{\bibfnamefont {M.~A.}\ \bibnamefont {Continentino}}, \bibinfo {author}
		{\bibfnamefont {A.}~\bibnamefont {Paduan-Filho}}, \bibinfo {author}
		{\bibfnamefont {B.}~\bibnamefont {Chiari}}, \ and\ \bibinfo {author}
		{\bibfnamefont {O.}~\bibnamefont {Piovesana}},\ }\href {\doibase
		10.1103/PhysRevLett.109.177206} {\bibfield  {journal} {\bibinfo  {journal}
			{Physical Review Letters}\ }\textbf {\bibinfo {volume} {109}},\ \bibinfo
		{pages} {177206} (\bibinfo {year} {2012})}\BibitemShut {NoStop}%
	\bibitem [{\citenamefont {Cvitani{\'c}}(2020)}]{CvitanicPhD}%
	\BibitemOpen
	\bibfield  {author} {\bibinfo {author} {\bibfnamefont {T.}~\bibnamefont
			{Cvitani{\'c}}},\ }\emph {\bibinfo {title} {Low dimensional quantum magnetism
			in {SeCuO$_3$} and {\textit{m}-NO$_2$PhBNO}}},\ \href@noop {} {\bibinfo
		{type} {Doctoral {Thesis}}},\ \bibinfo  {school} {University of Zagreb},
	\bibinfo {address} {Zagreb, Croatia} (\bibinfo {year} {2020})\BibitemShut
	{NoStop}%
	\bibitem [{\citenamefont {Sebastian}\ \emph {et~al.}(2005)\citenamefont
		{Sebastian}, \citenamefont {Sharma}, \citenamefont {Jaime}, \citenamefont
		{Harrison}, \citenamefont {Correa}, \citenamefont {Balicas}, \citenamefont
		{Kawashima}, \citenamefont {Batista},\ and\ \citenamefont
		{Fisher}}]{Sebastian}%
	\BibitemOpen
	\bibfield  {author} {\bibinfo {author} {\bibfnamefont {S.~E.}\ \bibnamefont
			{Sebastian}}, \bibinfo {author} {\bibfnamefont {P.~A.}\ \bibnamefont
			{Sharma}}, \bibinfo {author} {\bibfnamefont {M.}~\bibnamefont {Jaime}},
		\bibinfo {author} {\bibfnamefont {N.}~\bibnamefont {Harrison}}, \bibinfo
		{author} {\bibfnamefont {V.}~\bibnamefont {Correa}}, \bibinfo {author}
		{\bibfnamefont {L.}~\bibnamefont {Balicas}}, \bibinfo {author} {\bibfnamefont
			{N.}~\bibnamefont {Kawashima}}, \bibinfo {author} {\bibfnamefont {C.~D.}\
			\bibnamefont {Batista}}, \ and\ \bibinfo {author} {\bibfnamefont {I.~R.}\
			\bibnamefont {Fisher}},\ }\href {\doibase 10.1103/PhysRevB.72.100404}
	{\bibfield  {journal} {\bibinfo  {journal} {Physical Review B}\ }\textbf
		{\bibinfo {volume} {72}},\ \bibinfo {pages} {100404} (\bibinfo {year}
		{2005})}\BibitemShut {NoStop}%
	\bibitem [{\citenamefont {von Toussaint}(2011)}]{VonToussaint}%
	\BibitemOpen
	\bibfield  {author} {\bibinfo {author} {\bibfnamefont {U.}~\bibnamefont {von
				Toussaint}},\ }\href {\doibase 10.1103/RevModPhys.83.943} {\bibfield
		{journal} {\bibinfo  {journal} {Review of Modern Physics}\ }\textbf {\bibinfo
			{volume} {83}},\ \bibinfo {pages} {943} (\bibinfo {year} {2011})}\BibitemShut
	{NoStop}%
	\bibitem [{\citenamefont {Giamarchi}\ and\ \citenamefont
		{Tsvelik}(1999)}]{GiamarchiTsvelik}%
	\BibitemOpen
	\bibfield  {author} {\bibinfo {author} {\bibfnamefont {T.}~\bibnamefont
			{Giamarchi}}\ and\ \bibinfo {author} {\bibfnamefont {A.~M.}\ \bibnamefont
			{Tsvelik}},\ }\href {\doibase 10.1103/PhysRevB.59.11398} {\bibfield
		{journal} {\bibinfo  {journal} {Physical Review B}\ }\textbf {\bibinfo
			{volume} {59}},\ \bibinfo {pages} {11398} (\bibinfo {year}
		{1999})}\BibitemShut {NoStop}%
	\bibitem [{\citenamefont {Sakai}\ and\ \citenamefont
		{Takahashi}(1989)}]{Sakai1989}%
	\BibitemOpen
	\bibfield  {author} {\bibinfo {author} {\bibfnamefont {T.}~\bibnamefont
			{Sakai}}\ and\ \bibinfo {author} {\bibfnamefont {M.}~\bibnamefont
			{Takahashi}},\ }\href {\doibase 10.1143/JPSJ.58.3131} {\bibfield  {journal}
		{\bibinfo  {journal} {Journal of the Physical Society of Japan}\ }\textbf
		{\bibinfo {volume} {58}},\ \bibinfo {pages} {3131} (\bibinfo {year}
		{1989})}\BibitemShut {NoStop}%
	\bibitem [{\citenamefont {Wierschem}\ and\ \citenamefont
		{Sengupta}(2014{\natexlab{b}})}]{WierschemMPLB}%
	\BibitemOpen
	\bibfield  {author} {\bibinfo {author} {\bibfnamefont {K.}~\bibnamefont
			{Wierschem}}\ and\ \bibinfo {author} {\bibfnamefont {P.}~\bibnamefont
			{Sengupta}},\ }\href {\doibase 10.1142/S0217984914300178} {\bibfield
		{journal} {\bibinfo  {journal} {Modern Physics Letters B}\ }\textbf {\bibinfo
			{volume} {28}},\ \bibinfo {pages} {1430017} (\bibinfo {year}
		{2014}{\natexlab{b}})}\BibitemShut {NoStop}%
	\bibitem [{\citenamefont {Zapf}\ \emph {et~al.}(2007)\citenamefont {Zapf},
		\citenamefont {Correa}, \citenamefont {Batista}, \citenamefont {Murphy},
		\citenamefont {Palm}, \citenamefont {Jaime}, \citenamefont {Tozer},
		\citenamefont {Lacerda},\ and\ \citenamefont {Paduan-Filho}}]{zapf2007}%
	\BibitemOpen
	\bibfield  {author} {\bibinfo {author} {\bibfnamefont {V.~S.}\ \bibnamefont
			{Zapf}}, \bibinfo {author} {\bibfnamefont {V.~F.}\ \bibnamefont {Correa}},
		\bibinfo {author} {\bibfnamefont {C.~D.}\ \bibnamefont {Batista}}, \bibinfo
		{author} {\bibfnamefont {T.~P.}\ \bibnamefont {Murphy}}, \bibinfo {author}
		{\bibfnamefont {E.~D.}\ \bibnamefont {Palm}}, \bibinfo {author}
		{\bibfnamefont {M.}~\bibnamefont {Jaime}}, \bibinfo {author} {\bibfnamefont
			{S.}~\bibnamefont {Tozer}}, \bibinfo {author} {\bibfnamefont
			{A.}~\bibnamefont {Lacerda}}, \ and\ \bibinfo {author} {\bibfnamefont
			{A.}~\bibnamefont {Paduan-Filho}},\ }\href {\doibase 10.1063/1.2711612}
	{\bibfield  {journal} {\bibinfo  {journal} {Journal of Applied Physics}\
		}\textbf {\bibinfo {volume} {101}},\ \bibinfo {pages} {09E106} (\bibinfo
		{year} {2007})}\BibitemShut {NoStop}%
	\bibitem [{\citenamefont {Nomura}\ \emph {et~al.}(2022)\citenamefont {Nomura},
		\citenamefont {Matsuda}, \citenamefont {Ikeda}, \citenamefont {Kohama},
		\citenamefont {Tsuda}, \citenamefont {Amaya}, \citenamefont {Ono},\ and\
		\citenamefont {Hosokoshi}}]{Nomura}%
	\BibitemOpen
	\bibfield  {author} {\bibinfo {author} {\bibfnamefont {K.}~\bibnamefont
			{Nomura}}, \bibinfo {author} {\bibfnamefont {Y.~H.}\ \bibnamefont {Matsuda}},
		\bibinfo {author} {\bibfnamefont {A.}~\bibnamefont {Ikeda}}, \bibinfo
		{author} {\bibfnamefont {Y.}~\bibnamefont {Kohama}}, \bibinfo {author}
		{\bibfnamefont {H.}~\bibnamefont {Tsuda}}, \bibinfo {author} {\bibfnamefont
			{N.}~\bibnamefont {Amaya}}, \bibinfo {author} {\bibfnamefont
			{T.}~\bibnamefont {Ono}}, \ and\ \bibinfo {author} {\bibfnamefont
			{Y.}~\bibnamefont {Hosokoshi}},\ }\href {\doibase
		10.1103/PhysRevB.105.214430} {\bibfield  {journal} {\bibinfo  {journal}
			{Physical Review B}\ }\textbf {\bibinfo {volume} {105}},\ \bibinfo {pages}
		{214430} (\bibinfo {year} {2022})}\BibitemShut {NoStop}%
	\bibitem [{\citenamefont {Dupont}\ \emph {et~al.}(2018)\citenamefont {Dupont},
		\citenamefont {Capponi}, \citenamefont {Laflorencie},\ and\ \citenamefont
		{Orignac}}]{Dupont2018}%
	\BibitemOpen
	\bibfield  {author} {\bibinfo {author} {\bibfnamefont {M.}~\bibnamefont
			{Dupont}}, \bibinfo {author} {\bibfnamefont {S.}~\bibnamefont {Capponi}},
		\bibinfo {author} {\bibfnamefont {N.}~\bibnamefont {Laflorencie}}, \ and\
		\bibinfo {author} {\bibfnamefont {E.}~\bibnamefont {Orignac}},\ }\href
	{\doibase 10.1103/PhysRevB.98.094403} {\bibfield  {journal} {\bibinfo
			{journal} {Physical Review B}\ }\textbf {\bibinfo {volume} {98}},\ \bibinfo
		{pages} {094403} (\bibinfo {year} {2018})}\BibitemShut {NoStop}%
	\bibitem [{\citenamefont {Horvati\ifmmode~\acute{c}\else \'{c}\fi{}}\ \emph
		{et~al.}(2020)\citenamefont {Horvati\ifmmode~\acute{c}\else \'{c}\fi{}},
		\citenamefont {Klanj\ifmmode~\check{s}\else \v{s}\fi{}ek},\ and\
		\citenamefont {Orignac}}]{Horvatic2020}%
	\BibitemOpen
	\bibfield  {author} {\bibinfo {author} {\bibfnamefont {M.}~\bibnamefont
			{Horvati\ifmmode~\acute{c}\else \'{c}\fi{}}}, \bibinfo {author}
		{\bibfnamefont {M.}~\bibnamefont {Klanj\ifmmode~\check{s}\else
				\v{s}\fi{}ek}}, \ and\ \bibinfo {author} {\bibfnamefont {E.}~\bibnamefont
			{Orignac}},\ }\href {\doibase 10.1103/PhysRevB.101.220406} {\bibfield
		{journal} {\bibinfo  {journal} {Physical Review B}\ }\textbf {\bibinfo
			{volume} {101}},\ \bibinfo {pages} {220406} (\bibinfo {year}
		{2020})}\BibitemShut {NoStop}%
	\bibitem [{\citenamefont {Maeda}\ \emph {et~al.}(2007)\citenamefont {Maeda},
		\citenamefont {Hotta},\ and\ \citenamefont {Oshikawa}}]{Maeda2007}%
	\BibitemOpen
	\bibfield  {author} {\bibinfo {author} {\bibfnamefont {Y.}~\bibnamefont
			{Maeda}}, \bibinfo {author} {\bibfnamefont {C.}~\bibnamefont {Hotta}}, \ and\
		\bibinfo {author} {\bibfnamefont {M.}~\bibnamefont {Oshikawa}},\ }\href
	{\doibase 10.1103/PhysRevLett.99.057205} {\bibfield  {journal} {\bibinfo
			{journal} {Physical Review Letters}\ }\textbf {\bibinfo {volume} {99}},\
		\bibinfo {pages} {057205} (\bibinfo {year} {2007})}\BibitemShut {NoStop}%
	\bibitem [{\citenamefont {Orignac}\ \emph {et~al.}(2007)\citenamefont
		{Orignac}, \citenamefont {Citro},\ and\ \citenamefont
		{Giamarchi}}]{Orignac2007}%
	\BibitemOpen
	\bibfield  {author} {\bibinfo {author} {\bibfnamefont {E.}~\bibnamefont
			{Orignac}}, \bibinfo {author} {\bibfnamefont {R.}~\bibnamefont {Citro}}, \
		and\ \bibinfo {author} {\bibfnamefont {T.}~\bibnamefont {Giamarchi}},\ }\href
	{\doibase 10.1103/PhysRevB.75.140403} {\bibfield  {journal} {\bibinfo
			{journal} {Physical Review B}\ }\textbf {\bibinfo {volume} {75}},\ \bibinfo
		{pages} {140403} (\bibinfo {year} {2007})}\BibitemShut {NoStop}%
	\bibitem [{\citenamefont {Ranjith}\ \emph {et~al.}(2022)\citenamefont
		{Ranjith}, \citenamefont {Landolt}, \citenamefont {Raymond}, \citenamefont
		{Zheludev},\ and\ \citenamefont {Horvati\ifmmode~\acute{c}\else
			\'{c}\fi{}}}]{Ranjith2022}%
	\BibitemOpen
	\bibfield  {author} {\bibinfo {author} {\bibfnamefont {K.~M.}\ \bibnamefont
			{Ranjith}}, \bibinfo {author} {\bibfnamefont {F.}~\bibnamefont {Landolt}},
		\bibinfo {author} {\bibfnamefont {S.}~\bibnamefont {Raymond}}, \bibinfo
		{author} {\bibfnamefont {A.}~\bibnamefont {Zheludev}}, \ and\ \bibinfo
		{author} {\bibfnamefont {M.}~\bibnamefont {Horvati\ifmmode~\acute{c}\else
				\'{c}\fi{}}},\ }\href {\doibase 10.1103/PhysRevB.105.134429} {\bibfield
		{journal} {\bibinfo  {journal} {Physical Review B}\ }\textbf {\bibinfo
			{volume} {105}},\ \bibinfo {pages} {134429} (\bibinfo {year}
		{2022})}\BibitemShut {NoStop}%
	\bibitem [{\citenamefont {Daou}\ \emph {et~al.}(2010)\citenamefont {Daou},
		\citenamefont {Weickert}, \citenamefont {Nicklas}, \citenamefont {Steglich},
		\citenamefont {Haase},\ and\ \citenamefont {Doerr}}]{Daou2010}%
	\BibitemOpen
	\bibfield  {author} {\bibinfo {author} {\bibfnamefont {R.}~\bibnamefont
			{Daou}}, \bibinfo {author} {\bibfnamefont {F.}~\bibnamefont {Weickert}},
		\bibinfo {author} {\bibfnamefont {M.}~\bibnamefont {Nicklas}}, \bibinfo
		{author} {\bibfnamefont {F.}~\bibnamefont {Steglich}}, \bibinfo {author}
		{\bibfnamefont {A.}~\bibnamefont {Haase}}, \ and\ \bibinfo {author}
		{\bibfnamefont {M.}~\bibnamefont {Doerr}},\ }\href {\doibase
		10.1063/1.3356980} {\bibfield  {journal} {\bibinfo  {journal} {Review of
				Scientific Instruments}\ }\textbf {\bibinfo {volume} {81}},\ \bibinfo {pages}
		{033909} (\bibinfo {year} {2010})}\BibitemShut {NoStop}%
	\bibitem [{\citenamefont {Schollwöck}(2011)}]{Schollwock}%
	\BibitemOpen
	\bibfield  {author} {\bibinfo {author} {\bibfnamefont {U.}~\bibnamefont
			{Schollwöck}},\ }\href {\doibase https://doi.org/10.1016/j.aop.2010.09.012}
	{\bibfield  {journal} {\bibinfo  {journal} {Annals of Physics}\ }\textbf
		{\bibinfo {volume} {326}},\ \bibinfo {pages} {96} (\bibinfo {year} {2011})},\
	\bibinfo {note} {january 2011 Special Issue}\BibitemShut {NoStop}%
	\bibitem [{\citenamefont {Fishman}\ \emph {et~al.}(2022)\citenamefont
		{Fishman}, \citenamefont {White},\ and\ \citenamefont
		{Stoudenmire}}]{itensor}%
	\BibitemOpen
	\bibfield  {author} {\bibinfo {author} {\bibfnamefont {M.}~\bibnamefont
			{Fishman}}, \bibinfo {author} {\bibfnamefont {S.~R.}\ \bibnamefont {White}},
		\ and\ \bibinfo {author} {\bibfnamefont {E.~M.}\ \bibnamefont
			{Stoudenmire}},\ }\href {\doibase 10.21468/SciPostPhysCodeb.4} {\bibfield
		{journal} {\bibinfo  {journal} {SciPost Physics Codebases}\ ,\ \bibinfo
			{pages} {4}} (\bibinfo {year} {2022})}\BibitemShut {NoStop}%
	\bibitem [{\citenamefont {Blinder}\ \emph {et~al.}(2017)\citenamefont
		{Blinder}, \citenamefont {Dupont}, \citenamefont {Mukhopadhyay},
		\citenamefont {Grbić}, \citenamefont {Laflorencie}, \citenamefont {Capponi},
		\citenamefont {Mayaffre}, \citenamefont {Berthier}, \citenamefont
		{Paduan-Filho},\ and\ \citenamefont {Horvatić}}]{Blinder2017}%
	\BibitemOpen
	\bibfield  {author} {\bibinfo {author} {\bibfnamefont {R.}~\bibnamefont
			{Blinder}}, \bibinfo {author} {\bibfnamefont {M.}~\bibnamefont {Dupont}},
		\bibinfo {author} {\bibfnamefont {S.}~\bibnamefont {Mukhopadhyay}}, \bibinfo
		{author} {\bibfnamefont {M.~S.}\ \bibnamefont {Grbić}}, \bibinfo {author}
		{\bibfnamefont {N.}~\bibnamefont {Laflorencie}}, \bibinfo {author}
		{\bibfnamefont {S.}~\bibnamefont {Capponi}}, \bibinfo {author} {\bibfnamefont
			{H.}~\bibnamefont {Mayaffre}}, \bibinfo {author} {\bibfnamefont
			{C.}~\bibnamefont {Berthier}}, \bibinfo {author} {\bibfnamefont
			{A.}~\bibnamefont {Paduan-Filho}}, \ and\ \bibinfo {author} {\bibfnamefont
			{M.}~\bibnamefont {Horvatić}},\ }\href {\doibase 10.1103/PhysRevB.95.020404}
	{\bibfield  {journal} {\bibinfo  {journal} {Physical Review B}\ }\textbf
		{\bibinfo {volume} {95}},\ \bibinfo {pages} {1} (\bibinfo {year}
		{2017})}\BibitemShut {NoStop}%
	\bibitem [{\citenamefont {Sylju\aa{}sen}\ and\ \citenamefont
		{Sandvik}(2002)}]{Syljuasen}%
	\BibitemOpen
	\bibfield  {author} {\bibinfo {author} {\bibfnamefont {O.~F.}\ \bibnamefont
			{Sylju\aa{}sen}}\ and\ \bibinfo {author} {\bibfnamefont {A.~W.}\ \bibnamefont
			{Sandvik}},\ }\href {\doibase 10.1103/PhysRevE.66.046701} {\bibfield
		{journal} {\bibinfo  {journal} {Physical Review E}\ }\textbf {\bibinfo
			{volume} {66}},\ \bibinfo {pages} {046701} (\bibinfo {year}
		{2002})}\BibitemShut {NoStop}%
	\bibitem [{\citenamefont {Sandvik}(2010)}]{Sandvik}%
	\BibitemOpen
	\bibfield  {author} {\bibinfo {author} {\bibfnamefont {A.~W.}\ \bibnamefont
			{Sandvik}},\ }\href {\doibase 10.1063/1.3518900} {\bibfield  {journal}
		{\bibinfo  {journal} {AIP Conference Proceedings}\ }\textbf {\bibinfo
			{volume} {1297}},\ \bibinfo {pages} {135} (\bibinfo {year}
		{2010})}\BibitemShut {NoStop}%
\end{thebibliography}
%

\vspace{1 cm} \noindent{\bf Acknowledgments:} We acknowledge fruitful discussions with  P. Sengupta. \textbf{Funding:} M.S.G. and I.J. acknowledge the support of Croatian Science Foundation (HRZZ) under the project IP-2018-01-2970 and the support of project CeNIKS co-financed by the Croatian Government and the European Union through the European Regional Development Fund - Competitiveness and Cohesion Operational Programme (Grant No. KK.01.1.1.02.0013). M.D. completed his contribution to this work prior to joining Rigetti Computing. We acknowledge the support of LNCMI-CNRS and HLD-HZDR, members of the European Magnetic Field Laboratory (EMFL).

\end{document}